\documentclass[12pt,draftclsnofoot,onecolumn]{IEEEtran}
\IEEEoverridecommandlockouts

\usepackage{}
\usepackage{amssymb}
\usepackage{amsmath}

\usepackage{graphicx,bm,cite,epsfig,amssymb,amsmath,multirow,enumerate,amsthm}

\usepackage{paralist}
\usepackage[T1]{fontenc}
\usepackage[utf8]{inputenc}
\usepackage{authblk}
\usepackage{steinmetz}

\usepackage{psfrag}

\newtheorem{proposition}{Proposition}
\newtheorem{lemma}{Lemma}

\usepackage{algorithm, algorithmic}

\DeclareMathAlphabet{\mathbit}{OML}{cmr}{bx}{it}
\DeclareMathAlphabet{\mathsf}{OT1}{cmss}{m}{n}
\DeclareMathAlphabet{\mathTXf}{OT1}{cmss}{bx}{it}

\DeclareMathOperator{\rank}{rank}

\newcommand{\tr}{{\text{tr}}}

\newcommand{\He}{{{\mathrm{H}}}}

\theoremstyle{remark}
\newtheorem{remark}{Remark} 

\makeatletter
\renewcommand{\maketag@@@}[1]{\hbox{\m@th\normalsize\normalfont#1}}%
\makeatother

\begin{document}

\title{Joint Sensing and Reception Design of {S}{I}{M}{O} Hybrid Cognitive Radio Systems}

\author{Miltiades~C.~Filippou,~\IEEEmembership{Member,~IEEE,}
        George~A.~Ropokis,
				David~Gesbert,~\IEEEmembership{Fellow,~IEEE}
			  and 
				Tharmalingam~Ratnarajah,~\IEEEmembership{Senior~Member,~IEEE}
\thanks{This work is supported by the Seventh Framework Programme for Research of the European Commission under grant number ADEL-619647.~Part of this work has been presented in \cite{FilippouVTC2013}.~Dr.~Ropokis was funded by the MIMOCORD project.~The project is implemented within the framework of the Action ``Supporting Postdoctoral Researchers'' of the Operational Program ``Education and Lifelong Learning'' (Action's Beneficiary: General Secretariat for Research and Technology), and is co-financed by the European Social Fund (ESF) and the Greek State.}
\thanks{M.~C.~Filippou and T.~Ratnarajah are with the Institute for Digital Communications, University of Edinburgh, Edinburgh EH9 3FG, U.K. (e-mail: \{m.filippou, t.ratnarajah\}@ed.ac.uk).}
\thanks{G.A.~Ropokis is with Computer Technology Institute and Press ``Diophantus'', 26500, Rio-Patras, Greece (e-mail: ropokis@noa.gr).}
\thanks{D.~Gesbert is with EURECOM, Campus SophiaTech, 450 Route des Chappes, 06410, Biot, France (e-mail: gesbert@eurecom.fr).}
}

\IEEEpeerreviewmaketitle

\maketitle

\IEEEpeerreviewmaketitle

\begin{abstract}
 In this paper, the problem of joint design of Spectrum Sensing (SS) and receive beamforming (BF), with reference to a Cognitive Radio (CR) system, is considered.~The aim of the proposed design is the maximization of the achievable average uplink rate of a Secondary User (SU), subject to an outage-based Quality-of-Service (QoS) constraint for primary communication.~A hybrid CR system approach is studied, according to which, the system either operates as an interweave (i.e., opportunistic) or as an underlay (i.e., spectrum sharing) CR system, based on SS results.~A realistic Channel State Information (CSI) framework is assumed, according to which, the direct channel links are known by the multiple antenna receivers (RXs), while, merely statistical (covariance) information is available for the interference links.~A new, closed form approximation is derived for the outage probability of primary communication, and the problem of rate-optimal selection of SS parameters and receive beamformers is addressed for hybrid, interweave and underlay CR systems.~It is proven that our proposed system design outperforms both underlay and interweave CR systems for a range of system scenarios.
\end{abstract}

\IEEEpeerreviewmaketitle

\section{Introduction}
\label{sec:intro}
Spectrum scarcity, as it had been observed in 2002 by the Federal Communications Commission (FCC) \cite{federal2002spectrum}, constitutes a major drawback, in terms of facilitating wireless communications services.~To overcome such an obstacle, the notion of Cognitive Radio (CR) \cite{Mitola1999, Haykin2005, Biglieri2012, Goldsmith2009} was introduced, targeting at improving the information throughput by optimally exploiting the under-utilized spectrum.

In practice, two different categories of CR systems have been devised: \begin{inparaenum}[\itshape a\upshape)] \item \emph{Underlay} (or spectrum sharing) CR systems, where a Primary User (PU) allows the reuse of its spectrum by an unlicensed Secondary User (SU), \emph{provided} that the interference received by the PU will be such that an \emph{interference temperature} constraint will not be violated, and \item \emph{Interweave} (or opportunistic) CR systems, where the SU senses the spectrum environment and transmits at time intervals during which primary activity is not detected.\end{inparaenum}~As it has been explained in \cite{Filippou2015TWireless}, via an analytical comparative study, each of the described CR approaches is characterized by drawbacks of different kind.~For instance, the throughput performance of an interweave system is seriously affected by the quality of Spectrum Sensing (SS), while underlay CR systems, in turn, manipulate their transmission strategy according to a fixed interference temperature constraint, without exploiting the traffic pattern (or activity profile) of the PU.

With the aim of relaxing such inherent drawbacks, a \emph{hybrid} interweave/underlay CR approach has been investigated in the literature, in order to exploit the benefits of the two standard CR approaches.~However, the full potential of such a scheme, considering a realistic and practical system, from a Channel State Information (CSI) viewpoint, has not been studied so far to the best of our knowledge.~For instance, in works such as \cite{Jinhyung2010, Senthuran2012, Junni2013, Hojin2013}, hybrid CR systems are proposed, however, either no average rate-based performance analysis under channel fading is undertaken \cite{Jinhyung2010, Junni2013} or the unrealistic assumption of perfect SS is assumed \cite{Senthuran2012}.~In \cite{Hojin2013} a Single-Input-Single-Output (SISO) framework is investigated, thus, not being in accordance with today's most wireless systems, where multiple antennas are used at the Base Stations (BSs) and (possibly) at the mobile devices as well.~Furthermore, in \cite{Ropokis2015EW}, the problem of joint, optimal (in terms of average SU rate) SS and power policy design is investigated for a hybrid CR system in the uplink, however, assuming the existence of uncorrelated receive antennas and applying a Maximal Ratio Combining (MRC) receiver.~Also, in \cite{Filippou2015ICC}, the downlink of a Multiple-Input-Multiple-Output (MIMO) hybrid CR system is studied analytically and performance comparisons are made with the standard interweave and underlay CR systems.~Nonetheless, the existence of spatially uncorrelated antennas at the transmitters (TXs) is assumed, along with the application of a simple, truncated power allocation scheme, depending on an interference temperature threshold.~Moreover, in \cite{FilippouVTC2013}, the problem of optimal, in terms of the achievable average uplink rate, beamforming (BF) problem is presented and solved, focusing on the two-user, multiple-antenna interference channel, with combined instantaneous and statistical CSI.~However, the two systems are characterized by the same priority, thus, no solution for the equivalent CR system was provided.

Motivated by the above, in this paper we focus on the uplink of a hybrid interweave/underlay CR system.~The hybrid CR system operates either as an interweave or as an underlay CR system, based on the results of the SS procedure.~In such a setting, our contributions can be summarized as follows:
\begin{itemize}
\item Focusing on a spatially correlated fading channel model and assuming a combined CSI setting at the receivers (RXs) (CSIR), where direct links are known instantaneously and interference links are merely known based on their second order statistics, we derive new closed form approximations for the outage probability of primary communication, considering the hybrid CR system as well as the standard interweave and underlay CR systems.~Simulations show that the derived expressions approximate the actual outage probability sufficiently well.
\begin{itemize}
 \item Focusing on primary systems applying MRC receivers, the derived approximations are, to the best of our knowledge, the first appearing in the literature, that, unlike works such as \cite{Xiaodi2006}, also include additive noise and do not presume a specific relation between the covariance matrices of the desired and interfering channels.~Also, in contrast with \cite{Cui2006}, both the desired and the interference links are spatially correlated.
\end{itemize}
\item Having derived the expressions described above, and focusing on an interference-limited system, i.e., a system for which interference is the dominant source of signal degradation, as compared to noise, for the first time, we formulate and solve the problem of jointly determining \begin{inparaenum}[\itshape a\upshape)] \item the transmit power of the SU, \item the applied receive BF scheme, as well as \item the SS parameters, \end{inparaenum} such as to maximize the achievable average rate of the SU, subject to an outage-based constraint on primary communication.~The derived optimization framework is applied to all previously described CR system approaches, i.e., hybrid, as well as interweave/underlay.
\begin{itemize}
 \item The derived optimization framework can be applied for determining the transmit power and optimizing the BF and SS design for uplink communication of CR systems as well as for Licensed Shared Access (LSA) systems \cite{cept2014report, holdren2012realizing}, where the operation of a licensee user without violating the performance of an incumbent user, is crucial.
 \item Focusing, in particular, on the SS and receive BF optimization framework, we note that, to the best of our knowledge, the BF and SS problems are treated in a joint manner, for the first time.
\end{itemize}
\item The throughput performance of the optimized hybrid CR system is evaluated and compared to the performance achieved by the two optimized standard CR systems.~It turns out that the hybrid system outperforms the standard ones for the whole range of values of the investigated system design parameters, i.e., the outage constraint and the activity profile of the PU.~It is also shown that the performance of the hybrid CR system for low primary activity profiles, converges to the one achieved by the interweave system, while, for high primary activity profiles, the hybrid CR system behaves in a similar manner as the standard underlay one.
\end{itemize}

The following notations are adopted throughout the paper: all lower case boldface letters indicate vectors, whereas all upper case boldface letters denote matrices.~Superscript $(\cdot)^{\He}$ stands for Hermitian transpose, $\|\cdot\|$ denotes the Euclidean norm and $Pr(A)$ denotes the probability of event $A$.~Symbol ${[\mathbf{A}]}_{(p,q)}$ denotes the $(p,q)$-th element of matrix $\mathbf{A}$.~The all-zero vector of dimension $n \times 1$ is denoted as $\bm{0}_{n}$.~The identity matrix of dimension $n \times n$ is denoted as $\mathbf{I}_n$, whereas $\mathbb{E}_{|X}\{f(X,Y)\}$ symbolizes the conditional (with respect to Random Variable (RV) $X$) expectation of function $f(X,Y)$.~Also, $\tr(\mathbf{A})$, $\lambda_j(\mathbf{A})$ and $\rank(\mathbf{A})$ denote the trace, the $j$-th largest eigenvalue of square matrix $\mathbf{A}$ and its rank, respectively.~For a random vector $\bm{x}, \bm{x} \sim \mathcal{CN}(\boldsymbol{\mu}, \mathbf{\Sigma})$ denotes that $\bm{x}$ follows a Circularly Symmetric Complex Gaussian (CSCG) distribution, with mean $\boldsymbol{\mu}$ and covariance matrix $\mathbf{\Sigma}$.~Furthermore, $\exp(\cdot)$ and $\ln(\cdot)$ denote the exponential and logarithmic functions.~Additionally, $E_1(\cdot)$ represents the exponential integral function, as defined in \cite[eq. (5.1.1)]{Abramowitz} and $\mathcal{Q}(\cdot)$ represents the complementary Gaussian distribution function, as defined in \cite[eq (4.1)]{simon2005digital}.~Finally, $\gamma \approx 0.5772$ stands for the Euler-Mascheroni constant, as defined in \cite[eq. (4.1.32)]{Abramowitz}.

\section{System Model}

\subsection{Signal and channel model}
\begin{figure}[!ht]
  \centering
  \includegraphics[scale=0.58]{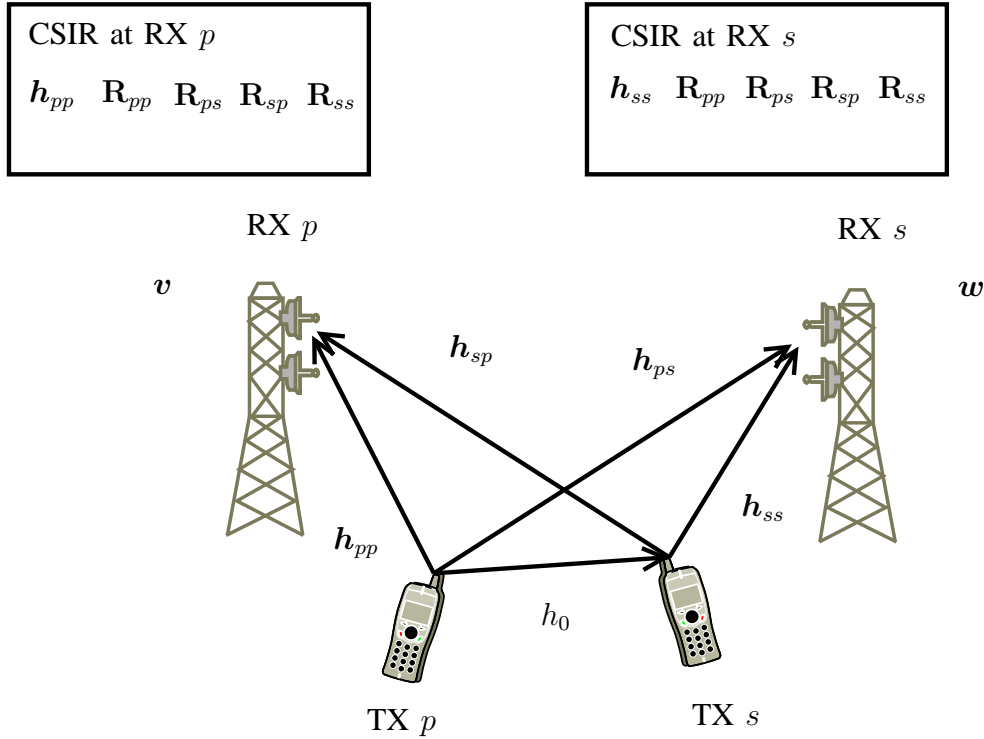}\\
	\vspace{-0.1in}
  \caption{The examined hybrid CR system.} \label{fig:SystemModel}
\end{figure}

The uplink of a CR system is considered, as shown in Fig.~\ref{fig:SystemModel}, which comprises of a single-antenna TX of a primary network, $\textrm{TX}~p$, that communicates with a multiple-antenna RX, $\textrm{RX}~p$.~It is assumed that the primary network is willing to share part of its spectral resources with a secondary network.~The latter is composed of a single-antenna TX, $\textrm{TX}~s$, communicating with a multiple antenna RX, $\textrm{RX}~s$.~In what follows, it is assumed that $\textrm{RX}~p$ and $\textrm{RX}~s$ are equipped with $M$ antennas, each.

The Single-Input-Multiple-Output (SIMO) channel between $\textrm{TX}~m$ and $\textrm{RX}~n$ is denoted as $\bm{h}_{mn} \in \mathbb{C}^{M \times 1}, \hspace{0.1in} m,n \in \{ p,s\}$ and the Rayleigh fading SISO channel between $\textrm{TX}~p$ and $\textrm{TX}~s$ is denoted as $h_0 \sim \mathcal{CN}(0, \sigma_0^2)$.~Also, the elements of channels $\bm{h}_{mn}, m, n \in \{p,s\}$, are spatially correlated, hence 
$\bm{h}_{mn} \sim \mathcal{CN}(\bm{0}_{M}, \mathbf{R}_{mn})$, with  $m,n \in \{p,s\}$ or
\begin{equation}
 \bm{h}_{mn} = \mathbf{R}_{mn}^{\frac{1}{2}} \bm{h}_{mn,w}, \hspace{0.1in} m,n \in \{p,s\},
\label{kronecker}
\end{equation}
where $\mathbf{R}_{mn}^{\frac{1}{2}}$ is the symmetric square root of covariance matrix $\mathbf{R}_{mn}$ of channel vector $\bm{h}_{mn}$ and $\bm{h}_{mn,w} \sim \mathcal{CN}(\bm{0}_{M}, \mathbf{I}_M)$.
\psfrag{v}{$\bm{v}$}
\psfrag{w}{$\bm{w}$}
\psfrag{h0}{$h_0$}
\psfrag{hpp}{$\bm{h}_{pp}$}
\psfrag{hss}{$\bm{h}_{ss}$}
\psfrag{hsp}{$\bm{h}_{sp}$}
\psfrag{hps}{$\bm{h}_{ps}$}
\psfrag{Rpp}{$\mathbf{R}_{pp}$}
\psfrag{Rps}{$\mathbf{R}_{ps}$}
\psfrag{Rsp}{$\mathbf{R}_{sp}$}
\psfrag{Rss}{$\mathbf{R}_{ss}$}
\psfrag{TXp}{$\textrm{TX}~p$}
\psfrag{TXs}{$\textrm{TX}~s$}
\psfrag{RXp}{$\textrm{RX}~p$}
\psfrag{RXs}{$\textrm{RX}~s$}
\psfrag{CSIR_at_RXp}{CSIR at $\textrm{RX}~p$}
\psfrag{CSIR_at_RXs}{CSIR at $\textrm{RX}~s$}

Regarding the availability of CSIR, a practical scenario is considered, according to which $\textrm{RX}~i, \hspace{0.1in} i \in \{p,s\}$, is aware of direct channel $\bm{h}_{ii}$, while it merely has statistical knowledge of the global uplink channel, in the form of covariance information.~Since such information is slow varying, it can be available at each of the RXs via a low capacity/high latency feedback link.~Such a CSIR formulation is chosen, because standard releases for 4G wireless systems require that a given terminal is allowed to report instantaneous CSI to its home BS, however, it cannot report such information to interfering BSs \cite{FilippouVTC2013}.

Since SS constitutes an essential feature of the investigated hybrid CR system, focusing on secondary communication, each Medium Access Control (MAC) frame of the SU, that has a duration of $T$ time units, consists of \begin{inparaenum}[\itshape a\upshape)] \item a SS subframe, the duration of which is $\tau$ time units, followed by \item a Data Transmission (DT) subframe, which lasts for the remaining $T-\tau$ time units.\end{inparaenum}~Concerning SS, we choose to apply Energy Detection (ED), since it is characterized by low implementation complexity and analytical expressions for the false alarm and detection probabilities.~Furthermore, it is assumed that the duration of each MAC frame is such that the involved wireless channels remain fixed.

In what follows, we describe the operation of the studied system, during the SS and DT subframes of each MAC frame.

\subsection{Description of SS phase}
Focusing on the application of ED for SS, it is assumed that $\textrm{TX}~s$ senses the wireless channel by sampling the received signal, with a sampling frequency denoted by $f_s$, therefore, SS is based on $N=\tau f_s$ samples.~We define event $\mathcal{H}_0$ as the one occurring when the primary system is idle, and its complementary event is denoted as $\mathcal{H}_1$.~The received signal at $\textrm{RX}~s$ for the $n$-th, $n=1,\ldots,N$, time instant is expressed, for each of hypotheses $\mathcal{H}_0$ and $\mathcal{H}_1$ as
\begin{equation}
y_s[n] =
\begin{cases}
\eta[n],& \textrm{if}\;\; \mathcal{H}_0 \\
 h_{0} \sqrt{P_p} x_p[n] + \eta[n], & \textrm{if}\;\; \mathcal{H}_1,
\end{cases}
\end{equation}
$1 \leq n \leq N$, where additive noise, $\eta[n]$ is a CSCG, independent, identically distributed  (i.i.d.) process with $\eta[n] \sim \mathcal{CN}(0, N_{0,0})$, $P_p$ denotes the fixed transmit power emitted by $\textrm{TX}~p$ and the information symbol $x_p[n]$ is selected from a CSCG codebook, i.e., $x_p[n] \sim \mathcal{CN}(0,1)$ and is independent of $\eta[n]$.~Under these assumptions, when ED is applied, based on a detection threshold, denoted by $\varepsilon, \hspace{0.1in} \varepsilon \geq 0$, a closed form expression describing the average (over channel fading) probability of false alarm, $\mathcal{P}_f(N,\varepsilon)$, as well as an approximation for the average probability of detection, $\mathcal{P}_d(N,\varepsilon)$, are derived in \cite{Filippou2015TWireless, Liang2008}.~These expressions are the following
\begin{equation}
  \mathcal{P}_f(N,\varepsilon) = \mathcal{Q}\bigg(\sqrt{N}\bigg(\frac{\varepsilon}{N_{0,0}} - 1\bigg)\bigg),
\label{P_f}
\end{equation}
and
\begin{equation}
  \mathcal{P}_{d}(N,\varepsilon) = \mathcal{Q}\bigg(\sqrt{N}\bigg(\frac{\varepsilon}{P_p \sigma_{0}^2 + N_{0,0}} - 1\bigg)\bigg).
\label{P_d}
\end{equation}

In the following, the DT phase for every MAC frame, is described.

\subsection{Description of DT phase}
Having described the SS procedure, we now focus on the DT subframe of the secondary MAC frame.~As explained earlier, the operation of the secondary network during the time intervals corresponding to these subframes depends on the obtained SS results.~Thus, for the description of the received signal during the DT subframe, one needs to discriminate between two SS decision cases.
\begin{itemize}
\item Case I:  Absence of primary transmissions is detected.~We denote this event as $\hat{\mathcal{H}}_0$.~Whenever such an event occurs, $\textrm{TX}~s$ transmits using a power level $P_s = P_0$.~On the other hand, $\textrm{RX}~s$ employs a receive BF vector $\bm{w} = \bm{w}_0(\bm{h}_{ss}) \in \mathbb{C}^{M \times 1}$ for the detection of the signal transmitted by the secondary terminal.
\item Case II: Presence of primary transmission is detected.~We denote this event as $\hat{\mathcal{H}}_1$.~Whenever $\hat{\mathcal{H}}_1$ occurs, $\textrm{TX}~s$ transmits using a power level, $P_s = P_1$.~In addition, $\textrm{RX}~s$ employs a BF vector, $\bm{w} = \bm{w}_1(\bm{h}_{ss}) \in \mathbb{C}^{M \times 1}$, that is designed taking into account the fact that primary activity has been detected.
\end{itemize}
In the following analysis, the achievable instantaneous rate at $\textrm{RX}~s$, regarding the investigated system model, is derived.
\subsection{Rate analysis of the secondary system}\label{rate_analysis_sec}
For the determination of the achievable instantaneous rate of the secondary system, the signal model at the RX side needs to be examined.~Using events $\mathcal{H}_0$ and $\mathcal{H}_1$, that were defined before, one can write the expression for the received signal reaching $\textrm{RX}~s$, after applying receive BF, given that event $\hat{\mathcal{H}}_k$ has occurred, as
\begin{equation}
 y_{k} = \bm{w}_k^{\He} \bm{h}_{ss} \sqrt{P_{k}} x_s 
+ c_k \bm{w}_k^{\He} \bm{h}_{ps} \sqrt{P_{p}}x_p + \bm{w}_k^H \bm{n}_{s},
\label{eq::cr_signal_model}
\end{equation}
$k \in \left\{0,1\right\}$, where $x_p, x_s$ denote the standard complex Gaussian signals transmitted by $\textrm{TX}~p$ and $\textrm{TX}~s$, respectively, and $\bm{n}_{s}$ denotes the noise at $\textrm{RX}~s$.~For the latter, it is assumed that $\bm{n}_s \sim \mathcal{CN} \left( \bm{0}_{M}, N_{0,s} \mathbf{I}_M\right)$.~Variable $c_k$ in \eqref{eq::cr_signal_model} is a discrete Bernoulli distributed RV, that models the presence of primary activity during secondary transmissions.~In more detail, RV $c_k$ is equal to $1$ with probability $\Pr \left( \mathcal{H}_1 | \hat{\mathcal{H}}_k\right)$.

Given \eqref{eq::cr_signal_model}, the achievable instantaneous secondary rate of the examined system model can be expressed as
\begin{equation}
\mathcal{R} = \mathcal{R}_0 + \mathcal{R}_1,
\end{equation}
where $\mathcal{R}_{k}, \hspace{0.1in} k \in \left\{ 0,1\right\}$, correspond to the rates achieved in Cases I and II, respectively.~More specifically, term $\mathcal{R}_k$ can be expressed as
\begin{equation}
\mathcal{R}_k {}={} 
\alpha_k
\log_2 \left( 1 + \frac{ \left|\bm{w}_{k}^{\He} \bm{h}_{ss} \right|^2 P_{k}}{N_{0,s}}\right)
+ 
\beta_k
\log_2 \left( 
1 + \frac{ \left|\bm{w}_{k}^{\He} \bm{h}_{ss} \right|^2 P_{k}}
{N_{0,s} + \left|\bm{w}_{k}^{\He} \bm{h}_{ps} \right|^2 P_p}
\right),
\label{eq::rate_i}
\end{equation}
where \eqref{eq::rate_i} holds under the assumption that $\left\| \bm{w}_{k}\right\| = 1$.~Coefficients $\alpha_k, \beta_k$ in \eqref{eq::rate_i} are defined as
\begin{equation}
\begin{split}
{}&{}\alpha_0 = \frac{T - \tau}{T} \mathcal{P}_0 \left( 1 - \mathcal{P}_{f}\right), \hspace{0.2in}
\beta_0 = \frac{T - \tau}{T} \mathcal{P}_{1} \left( 1 - \mathcal{P}_d\right), \text{and}\\&
\alpha_1 = \frac{T - \tau}{T} \mathcal{P}_0 \mathcal{P}_{f}, \hspace{0.2in}
\beta_1 = \frac{T - \tau}{T} \mathcal{P}_1 \mathcal{P}_{d},
\end{split}
\label{alphas_betas}
\end{equation}
where $\mathcal{P}_0 = Pr(\mathcal{H}_0)$ and  $\mathcal{P}_1 = 1 - \mathcal{P}_0$.~Having presented the signal model and the achievable instantaneous rate for the secondary system, in the following section we focus on the received signal model for the primary system.
\subsection{Primary system operation mode}
Based on  the described operation mode of the secondary network, one can write the expression describing the received signal at $\textrm{RX}~p$, after applying receive BF, provided that primary transmission takes place, as
\begin{equation}
z_k = \bm{v}^{\He} \bm{h}_{pp} \sqrt{P_p} x_p + \bm{v}^{\He} \bm{h}_{sp} \sqrt{P_k} x_s + \bm{v}^{\He} \bm{n}_p, \hspace{0.1in} \textrm{if} \hspace{0.1in} \hat{\mathcal{H}}_k, \hspace{0.1in} k \in \{0, 1\},
\end{equation}
where $\bm{n}_{p}$ stands for the additive CSCG noise received by $\textrm{RX}~p$ and $\bm{v}$ represents the applied receive BF vector at $\textrm{RX}~p$, which is assumed to be a vector based on the MRC BF solution, thus $\bm{v} = \tilde{\bm{h}}_{pp} = \frac{\bm{h}_{pp}}{\|\bm{h}_{pp}\|}$.~In our analysis, we assume that $\bm{n}_{p} \sim \mathcal{CN} \left( \bm{0}_{M}, N_{0,p} \mathbf{I}_M\right)$.

Based on the above described system model, in the following sections we initially investigate the QoS, quantified by means of the targeted outage probability, that is achieved for primary communication, as well as the achievable average rate of secondary communication.~Following that, we formulate the problem of optimal SS and reception for the secondary RX, with emphasis on the maximization of its achievable average rate, given QoS-based constraints, related to the operation of the primary system.
\section{Preliminary Analytical Results}
In this section, closed form approximations describing the outage probability of primary communication, as well as the achievable average rate of secondary communication, are derived, focusing on a combined CSIR assumption.~According to this assumption, the direct channel links can be instantaneously available by the RXs, whereas the interference links are merely known by their channel covariance matrices.~In what follows, an approximation of the outage probability of the primary RX is derived in closed form.
\subsection{Outage probability of primary communication}
An outage event is declared at $\textrm{RX}~p$, when, given that primary transmissions take place, the Signal-to-Interference-plus-Noise Ratio (SINR) measured at $\textrm{RX}~p$ is below a threshold, denoted by $\gamma_0$.~In the following proposition, an analytical approximation for the outage probability experienced at $\textrm{RX}~p$, is derived.

\begin{proposition}\label{Outage_prob}
 The outage probability of primary communication, for a hybrid SIMO CR system can be approximated as

\begin{equation}
 \mathcal{P}_{\textrm{out}} \approx (1 - \mathcal{P}_d) \mathcal{F}(P_0) + \mathcal{P}_d \mathcal{F}(P_1), 
\label{P_out}
\end{equation}
where function $\mathcal{F}(x)$ is given by
\begin{equation}
 \mathcal{F}(x) = \frac{\exp\left(\frac{N_{0,p}}{x \bar{\lambda}}\right)}{\prod_{j=1}^M \lambda_j(\mathbf{R}_{pp})} \sum_{j=1}^M \frac{\frac{\lambda_j(\mathbf{R}_{pp}) \gamma_0 \bar{\lambda} x}{P_p \lambda_j(\mathbf{R}_{pp}) + \gamma_0 x \bar{\lambda}}}{\prod_{k=1, k \neq j}^M \left(\frac{1}{\lambda_k(\mathbf{R}_{pp})} - \frac{1}{\lambda_j(\mathbf{R}_{pp})}\right)},
\label{function_f}
\end{equation}
and $\bar{\lambda} = \mathbb{E}\left\{\frac{\bm{h}_{pp}^{\He} \mathbf{R}_{sp} \bm{h}_{pp}}{{\|\bm{h}_{pp}\|}^2}\right\}$ can be found in closed form by applying \cite[Lemma 3]{FilippouStatCoord2015}.

\begin{proof}
 The proof is included in Appendix \ref{proof_Outage_prob}.
\end{proof}
\end{proposition}
Having derived an analytical approximation describing the outage probability of primary communication, in what follows, we derive a lower bound for the average rate of secondary communication, given the instantaneous knowledge of direct channel $\bm{h}_{ss}$ at $\textrm{RX}~s$.

\subsection{Achievable average rate of secondary communication}
Given the analysis described in \ref{rate_analysis_sec}, the achievable ergodic rate experienced at $\textrm{RX}~s$, conditioned on the knowledge of channel $\bm{h}_{ss}$, is given by the following expression
\begin{equation}
 \mathbb{E}_{|\bm{h}_{ss}}\left\{\mathcal{R}\right\} = \mathbb{E}_{|\bm{h}_{ss}}\left\{\mathcal{R}_0\right\} + \mathbb{E}_{|\bm{h}_{ss}}\left\{\mathcal{R}_1\right\},
\label{SU_rate_total}
\end{equation}
where the occurrence of event $\hat{\mathcal{H}}_k, k=0,1$ is considered for each term $\mathbb{E}_{|\bm{h}_{ss}}\left\{\mathcal{R}_k\right\}$.~Hence, this leads us to the following analysis:
\subsubsection{Occurrence of event $\hat{\mathcal{H}}_0$}
When no primary activity is detected, as a result of SS, $\textrm{RX}~s$ applies the MRC BF solution such as to maximize the direct signal power, i.e., $\bm{w}_{0} = \tilde{\bm{h}}_{ss} = \frac{\bm{h}_{ss}}{\|\bm{h}_{ss}\|}$.~Also, $\textrm{TX}~s$, in its turn, can transmit with its full available instantaneous power, denoted by $P_{\textrm{peak}}$, i.e., $P_0 = P_{\textrm{peak}}$.~In this case, the average secondary rate, conditioned on the knowledge of channel $\bm{h}_{ss}$ at $\textrm{RX}~s$, is given by the lemma that follows.
\begin{lemma}\label{SU_rate_interweave}
The achievable average rate of secondary communication, conditioned on the instantaneous knowledge of channel $\bm{h}_{ss}$ at $\textrm{RX}~s$, when event $\hat{\mathcal{H}}_0$ occurs, is characterized by the following lower bound
\begin{equation}
 \mathbb{E}_{|\bm{h}_{ss}}\left\{\mathcal{R}_0\right\} \geq \mathcal{C}_0,
\label{SU_rate_inter_final}
\end{equation}
where
\begin{equation}
 \mathcal{C}_0 = \frac{\alpha_0}{\ln(2)} \mathcal{C}_{0,0} + \frac{\beta_0}{\ln(2)} \mathcal{C}_{0,1},
\label{C_0}
\end{equation}
and
\begin{equation}
 \mathcal{C}_{0,0} = \ln\left(1 + \frac{P_{\textrm{peak}} \|\bm{h}_{ss}\|^2}{N_{0,s}}\right), \hspace{0.1in} \mathcal{C}_{0,1} = \ln\left(1 + \frac{P_{\textrm{peak}} \|\bm{h}_{ss}\|^2}{N_{0,s} + P_p \frac{\bm{h}_{ss}^{\He} \mathbf{R}_{ps} \bm{h}_{ss}}{\|\bm{h}_{ss}\|^2}}\right).
\label{C_0_0_and_C_01}
\end{equation}
\begin{proof}
 The proof is included in Appendix \ref{proof_SU_rate_interweave}.
\end{proof}
\end{lemma}

\subsubsection{Occurrence of event $\hat{\mathcal{H}}_1$}
When event $\hat{\mathcal{H}}_1$ occurs, i.e., when the secondary system adopts the underlay CR approach, the receive BF vector $\bm{w}_{1}$ and the transmit power $P_1$, are system parameters which need to be designed.
\begin{lemma}\label{SU_rate_underlay}
The achievable average rate of secondary communication, conditioned on the instantaneous knowledge of channel $\bm{h}_{ss}$ at $\textrm{RX}~s$, when event $\hat{\mathcal{H}}_1$ occurs, is given by the following expression
\begin{equation}
 \mathbb{E}_{|\bm{h}_{ss}}\left\{\mathcal{R}_1\right\} = \frac{\alpha_1}{\ln(2)} \mathcal{C}_{1,0}  + \frac{\beta_1}{\ln(2)} \mathcal{C}_{1,1},
\label{SU_rate_under_final}
\end{equation}
where
\begin{equation}
 \mathcal{C}_{1,0} = \ln\left(1 + \frac{P_1}{N_{0,s}} |\bm{w}_{1}^{\He} \bm{h}_{ss}|^2 \right),
\label{C_1_0}
\end{equation}
and
\small
\begin{equation}
\begin{aligned}
 \mathcal{C}_{1,1} &= \ln\left(1 + \frac{P_1}{N_{0,s}} |\bm{w}_{1}^{\He} \bm{h}_{ss}|^2 \right) + \exp\left(\frac{\bm{w}_{1}^{\He} \left(\mathbf{I}_M + \frac{P_1}{N_{0,s}} \bm{h}_{ss} \bm{h}_{ss}^{\He}\right) \bm{w}_{1}}{\bm{w}_{1}^{\He} \rho_{\textrm{inr},s} \mathbf{R}_{ps} \bm{w}_{1}}\right) E_1\left(\frac{\bm{w}_{1}^{\He} \left(\mathbf{I}_M + \frac{P_1}{N_{0,s}} \bm{h}_{ss} \bm{h}_{ss}^{\He}\right) \bm{w}_{1}}{\bm{w}_{1}^{\He} \rho_{\textrm{inr},s} \mathbf{R}_{ps} \bm{w}_{1}}\right) \\
&- \exp\left(\frac{1}{\bm{w}_{1}^{\He} \rho_{\textrm{inr},s} \mathbf{R}_{ps} \bm{w}_{1}}\right) E_1\left(\frac{1}{\bm{w}_{1}^{\He} \rho_{\textrm{inr},s} \mathbf{R}_{ps} \bm{w}_{1}}\right),
\end{aligned}
\label{C_1_1}
\end{equation}
\normalsize
where $\rho_{\textrm{inr},s} = \frac{P_p}{N_{0,s}}$ is the system Interference-to-Noise Ratio (INR), received at $\textrm{RX}~s$, due to primary transmission.
\begin{proof}
 The proof is included in Appendix \ref{proof_SU_rate_underlay}.
\end{proof}
\end{lemma}
In the section that follows, an optimization problem is formulated, according to which the SS parameters are jointly optimized with the receive BF scheme applied at $\textrm{RX}~s$, with the aim of maximizing the conditional (for a given, known instant of channel $\bm{h}_{ss}$ at $\textrm{RX}~s$) average rate of secondary communication, subject to constraints, which are destined to protect primary transmissions.

At this point, it should be noted that, in the remainder of the paper, we will focus on an interference-limited CR system, i.e., a system in which interference is the main source of signal degradation, as compared to noise \cite{FilippouVTC2013}\footnote{Such an assumption is realistic for a CR scenario, as the secondary system can be in the vicinity of the primary, following a non-cooperative behavior.}.~With such an assumption, it holds that $\bm{w}_1^{\He} \rho_{\textrm{inr,s}} \mathbf{R}_{ps} \bm{w}_1 \gg N_{0,s}$, hence, assuming that $N_{0,s} = 1$, the last term of expression \eqref{C_1_1} asymptotically converges to \cite[eq. (5.1.11)]{Abramowitz}
\begin{equation}
 \exp\left(\frac{1}{\bm{w}_{1}^{\He} \rho_{\textrm{inr},s} \mathbf{R}_{ps} \bm{w}_{1}}\right) E_1\left(\frac{1}{\bm{w}_{1}^{\He} \rho_{\textrm{inr},s} \mathbf{R}_{ps} \bm{w}_{1}}\right) \xrightarrow{\bm{w}_{1}^{\He} \rho_{\textrm{inr},s} \mathbf{R}_{ps} \bm{w}_{1} \gg 1} - \gamma + \ln(\bm{w}_{1}^{\He} \rho_{\textrm{inr},s} \mathbf{R}_{ps} \bm{w}_{1}).
\label{approx_rate_high_INR}
\end{equation}
Consequently, incorporating the high INR assumption, the expectation $\mathbb{E}_{|\bm{h}_{ss}}\left\{\mathcal{R}_1\right\}$ becomes as follows
\begin{equation}
 \mathbb{E}_{|\bm{h}_{ss}}\left\{\mathcal{R}_1\right\} \xrightarrow{high \hspace{0.05in} INR} \frac{\alpha_1}{\ln(2)} \mathcal{D}_{1,0} + \frac{\beta_1}{\ln(2)} \mathcal{D}_{1,1}.
\label{approx_high_INR}
\end{equation}
Quantities $\mathcal{D}_{1,0}$ and $\mathcal{D}_{1,1}$ are given by
\begin{equation}
 \mathcal{D}_{1,0} = \mathcal{C}_{1,0},
\label{D_1_0}
\end{equation}
and 
\begin{equation}
\begin{aligned}
 \mathcal{D}_{1,1} &= \ln\left(\frac{\bm{w}_{1}^{\He} \left(\mathbf{I}_M + \frac{P_1}{N_{0,s}} \bm{h}_{ss} \bm{h}_{ss}^{\He}\right) \bm{w}_{1}}{\bm{w}_{1}^{\He} \rho_{\textrm{inr},s} \mathbf{R}_{ps} \bm{w}_{1}}\right) \\
&+ \exp\left(\frac{\bm{w}_{1}^{\He} \left(\mathbf{I}_M + \frac{P_1}{N_{0,s}} \bm{h}_{ss} \bm{h}_{ss}^{\He}\right) \bm{w}_{1}}{\bm{w}_{1}^{\He} \rho_{\textrm{inr},s} \mathbf{R}_{ps} \bm{w}_{1}}\right) E_1\left(\frac{\bm{w}_{1}^{\He} \left(\mathbf{I}_M + \frac{P_1}{N_{0,s}} \bm{h}_{ss} \bm{h}_{ss}^{\He}\right) \bm{w}_{1}}{\bm{w}_{1}^{\He} \rho_{\textrm{inr},s} \mathbf{R}_{ps} \bm{w}_{1}}\right) + \gamma,
\end{aligned}
\label{D_1_1}
\end{equation}
respectively.

\section{Problem Formulation}
Having derived a lower bound for the average rate of secondary communication as well as a closed form approximation for the outage probability of primary communication, an optimization problem can be formulated, the solution of which will lead to a rate-optimal scheme of SS and receive BF, with respect to the secondary system, given an outage-based constraint, which aims at protecting primary communication from harmful interference.~More specifically, the parameters that need to be optimized in such a direction, are: \begin{inparaenum}[\itshape a\upshape)] \item the SS design parameters, i.e., the sensing time, $\tau$ as well as the ED threshold, $\varepsilon$ and \item the receive BF vector, $\bm{w}_1$, applied at $\textrm{RX}~s$, when event $\hat{\mathcal{H}}_1$ occurs.\end{inparaenum}~Hence, the investigated optimization problem can be mathematically expressed as follows
\begin{equation}
\begin{aligned}
  & \underset{\bm{w}_{1} \in \mathbb{C}^{M \times 1}, \tau, \varepsilon, P_1}{\text{maximize}}
	& & \mathbb{E}_{|\bm{h}_{ss}}\left\{\mathcal{R}\right\} \\
	& \text{subject to} 
	& & \mathcal{P}_{\textrm{out}} \leq \tilde{\mathcal{P}}_{\textrm{out}}, \hspace{0.1in} \mathcal{P}_d = \tilde{\mathcal{P}}_d, \hspace{0.1in} \|\bm{w}_{1}\| = 1, \\
  &&& 0 < P_1 \leq P_{\textrm{peak}}, \hspace{0.1in} 0 < \tau \leq T, \hspace{0.1in} \varepsilon \geq 0,
\end{aligned}
\tag{P1}
\label{P1}
\end{equation}
where, $\tilde{\mathcal{P}}_{\textrm{out}}$ is the predetermined outage-based constraint, and $\tilde{\mathcal{P}}_d$ is a targeted average detection probability for the SS algorithm.

Solving problem \eqref{P1} proves to be complicated.~Thus, we propose to determine the rate-optimal BF and SS parameters by solving a simpler optimization problem.~The objective function of the new optimization problem is a lower bound of the average rate of $\textrm{RX}~s$, which is easier to manipulate.~This lower bound is: $\mathcal{C} = \mathcal{C}_0 + \mathbb{E}_{|\bm{h}_{ss}}\{\mathcal{R}_1\}$.~As a result, the optimization problem to be solved is the following
\begin{equation}
\begin{aligned}
  & \underset{\bm{w}_{1} \in \mathbb{C}^{M \times 1}, \tau, \varepsilon, P_1}{\text{maximize}}
	& & \mathcal{C} \\
	& \text{subject to} 
	& & \mathcal{P}_{\textrm{out}} \leq \tilde{\mathcal{P}}_{\textrm{out}}, \hspace{0.1in} \mathcal{P}_d = \tilde{\mathcal{P}}_d, \hspace{0.1in} \|\bm{w}_{1}\| = 1, \\
  &&& 0 < P_1 \leq P_{\textrm{peak}}, \hspace{0.1in} 0 < \tau \leq T, \hspace{0.1in} \varepsilon \geq 0.
\end{aligned}
\tag{P2}
\label{P2}
\end{equation}

At this stage, we choose to divide optimization problem \eqref{P2} into a number of sub-problems.~Focusing on each sub-problem, one parameter is optimized for given values of the remaining design parameters, which fulfill the constraints.
\section{Solving the Optimization Problem}
\subsection{Determining the transmit power of $\textrm{TX}~s$}\label{Opt_PP}
Clearly, the transmit power level, $P_1$, that maximizes the average rate of the SU, will be satisfying the outage constraint determined by the primary system, with equality. Hence, one needs to solve the following equation
\begin{equation}
 \left(1-\tilde{\mathcal{P}}_d\right) \mathcal{F}\left(P_{\textrm{peak}}\right) + \tilde{\mathcal{P}}_d \mathcal{F}\left(P_{1,\textrm{root}}\right) = \tilde{\mathcal{P}}_{\textrm{out}},
\label{PP_1}
\end{equation} 
with respect to parameter $P_{1,\textrm{root}}$.~As a result, the following equation is obtained
\begin{equation}
\begin{aligned}
 &P_{1,\textrm{root}} = \mathcal{F}^{-1}\left(y_0\right), \text{where} \hspace{0.1in} y_0 = \frac{\tilde{\mathcal{P}}_{\textrm{out}} - (1 - \tilde{\mathcal{P}}_d) \mathcal{F}\left(P_{\textrm{peak}}\right)}{\tilde{\mathcal{P}}_d}.
\end{aligned}
\label{PP_2}
\end{equation} 
The inversion of function $\mathcal{F}(\cdot)$ leads to a non-closed form expression, thus, a root finding method can be applied in terms of solving equation $\mathcal{F}(P_{1,\textrm{root}}) - y_0 = 0$, with respect to $P_{1,\textrm{root}} > 0$.~Hence, taking into consideration the peak power constraint at $\textrm{TX}~s$, the solution becomes
\begin{equation}
 P_1^{*} = \min\left\{P_{1,\textrm{root}}, P_{\textrm{peak}}\right\}.
\label{PP_3}
\end{equation} 
In the section that follows, an iterative scheme of jointly optimizing the receive BF vector and the SS parameters, is thoroughly described.

\subsection{Jointly optimizing the receive BF vector and the SS parameters}\label{Opt_BF}
Having determined the applied transmit power at $\textrm{TX}~s$, $P_1^{*}$, which satisfies the outage probability constraint of problem \eqref{P2} with equality, the resulting optimization problem that needs to be solved is the following
\begin{equation}
\begin{aligned}
  & \underset{\bm{w}_{1} \in \mathbb{C}^{M \times 1}, \tau, \varepsilon}{\text{maximize}}
	& & \mathcal{C} \\
	& \text{subject to} 
	& &\mathcal{P}_d = \tilde{\mathcal{P}}_d, \hspace{0.1in} \|\bm{w}_{1}\| = 1, \\
  &&& 0 < \tau \leq T, \hspace{0.1in} \varepsilon \geq 0.
\end{aligned}
\tag{P3}
\label{P3}
\end{equation}
One can write the objective function of optimization problem \eqref{P3} as follows
\begin{equation}
 \mathcal{C}(\bm{w}_1, \tau, \varepsilon, P_1^{*}) = \mathcal{C}_0(\tau, \varepsilon) + \mathbb{E}_{|\bm{h}_{ss}}\left\{\mathcal{R}_1(\bm{w}_1, \tau, \varepsilon, P_1^{*})\right\}.
\label{obj_P3} 
\end{equation}
In order to approximate the solution to this problem, we propose to use an iterative procedure based on alternating optimization of the SS parameters and the receive BF vector.~Following such an approach requires solving the following two sub-problems.

\subsubsection{Optimizing the SS parameters for a given BF vector}\label{optim_SS_fix_BF}
We start with fixing the receive BF vector to be an arbitrary unit-norm vector, i.e., $\bm{w}_1 = \hat{\bm{w}}_1, \hspace{0.1in} \|\hat{\bm{w}}_1\| = 1$.~As a consequence, the resulting objective function of problem \eqref{P3} is only a function of SS parameters $\tau$ and $\varepsilon$, i.e., $\mathcal{C} = \mathcal{C}(\hat{\bm{w}}_1, \tau, \varepsilon, P_1^{*})$.~As a result, optimization problem \eqref{P3} becomes
\begin{equation}
\begin{aligned}
  & \underset{\tau, \varepsilon}{\text{maximize}}
	& & \tilde{\mathcal{C}}(\tau, \varepsilon) = \alpha_0(\tau, \varepsilon) \mathcal{C}_{0,0} + \beta_0(\tau, \varepsilon) \mathcal{C}_{0,1} + \alpha_1(\tau, \varepsilon) \hat{\mathcal{D}}_{1,0} + \beta_1(\tau, \varepsilon) \hat{\mathcal{D}}_{1,1}  \\
	& \text{subject to} 
	& & \mathcal{P}_d(\tau, \varepsilon) = \tilde{\mathcal{P}}_d, \hspace{0.1in} 0 < \tau \leq T, \hspace{0.1in} \varepsilon \geq 0,
\end{aligned}
\tag{P4}
\label{P4}
\end{equation}
where terms $\hat{\mathcal{D}}_{1,0}$ and $\hat{\mathcal{D}}_{1,1}$ are given by equations \eqref{D_1_0} and \eqref{D_1_1}, respectively, with $\bm{w}_1 = \hat{\bm{w}}_1$ and $P_1 = P_1^{*}$.
Exploiting the equality constraint for the average detection probability, along with expression \eqref{P_d}, one can express the ED threshold, $\varepsilon$, as a function of sensing time, $\tau$.~This expression is the following
\begin{equation}
 \varepsilon(\tau) = N_{0,0}\left(1 + \frac{P_p}{N_{0,0}} \sigma_{0}^2\right) \left(\frac{\mathcal{Q}^{-1}(\tilde{\mathcal{P}}_d)}{\sqrt{\tau f_s}} + 1\right).
\label{SS_1}
\end{equation} 
Substituting \eqref{SS_1} to the objective function of \eqref{P4}, the following lemma can be proved, which is useful for the solution of \eqref{P4}.
\begin{lemma}\label{Concave_SS}
Function $\tilde{\mathcal{C}}(\tau, \varepsilon(\tau))$ which is obtained after substituting \eqref{SS_1} to the objective function of \eqref{P4}, is a concave function for every $\tau \in (0,T]$.
\begin{proof}
 The proof is included in Appendix \ref{proof_Concave_SS}.
\end{proof}
\end{lemma}
Since the resulting optimization problem is a convex problem, any convex optimization algorithm can be applied (i.e., a gradient ascent-based algorithm), with the aim of finding the rate-optimal values $\tau^{*}$ as well as $\varepsilon^{*}$ (through \eqref{SS_1}), for the given receive BF vector, $\hat{\bm{w}}_1$.

\subsubsection{Optimizing the receive BF scheme for fixed SS parameters}\label{opt_BF_fixed_SS}
The problem of designing receive BF vector $\bm{w}_1$, such as to maximize the objective function of problem \eqref{P3}, for given SS parameters that satisfy the detection probability constraint, is equivalently expressed as follows
\begin{equation}
\begin{aligned}
  & \underset{\bm{w}_{1} \in \mathbb{C}^{M \times 1}}{\text{maximize}}
	& & \mathbb{E}_{|\bm{h}_{ss}}\left\{\mathcal{R}_1(\bm{w}_1, \hat{\tau}, \hat{\varepsilon}, P_1^{*})\right\} \\
	& \text{subject to} 
	& &\mathcal{P}_d(\hat{\tau}, \hat{\varepsilon}) = \tilde{\mathcal{P}}_d, \hspace{0.1in} \|\bm{w}_{1}\| = 1, \\
\end{aligned}
\tag{P5}
\label{P5}
\end{equation}
where $\hat{\tau} \in (0,T]$ and $\hat{\varepsilon} \geq 0$.~The objective function of problem \eqref{P5} is given by \eqref{approx_high_INR}, with $P_1 = P_1^{*}$, $\alpha_1 = \alpha_1(\hat{\tau}, \hat{\varepsilon}) = \hat{\alpha}_1$ and $\beta_1 = \beta_1(\hat{\tau}, \hat{\varepsilon}) = \hat{\beta}_1$.
Consequently, incorporating the high INR assumption, the objective function of the receive BF problem becomes as follows
\begin{equation}
\begin{aligned}
 &\mathbb{E}_{|\bm{h}_{ss}}\left\{\mathcal{R}_1\right\} \xrightarrow{high \hspace{0.05in} INR} \frac{\hat{\alpha}_1}{\ln(2)} \ln\left(\bm{w}_{1}^{\He} \mathbf{H}_{\textrm{eff}} \bm{w}_{1}\right) + \frac{\hat{\beta}_1}{\ln(2)} \Bigg(\ln\left(\frac{\bm{w}_{1}^{\He} \mathbf{H}_{\textrm{eff}} \bm{w}_{1}}{\bm{w}_{1}^{\He} \mathbf{R}_{\textrm{eff}} \bm{w}_{1}} \right) \\
&+ \exp\left(\frac{\bm{w}_{1}^{\He} \mathbf{H}_{\textrm{eff}} \bm{w}_{1}}{\bm{w}_{1}^{\He} \mathbf{R}_{\textrm{eff}} \bm{w}_{1}}\right) E_1\left(\frac{\bm{w}_{1}^{\He} \mathbf{H}_{\textrm{eff}} \bm{w}_{1}}{\bm{w}_{1}^{\He} \mathbf{R}_{\textrm{eff}} \bm{w}_{1}}\right) + \gamma \Bigg),
\end{aligned}
\label{SU_rate_high_INR}
\end{equation}
where $\mathbf{H}_{\textrm{eff}} = \mathbf{I}_M + \frac{P_1^{*}}{N_{0,s}} \bm{h}_{ss} \bm{h}_{ss}^{\He}$ and $\mathbf{R}_{\textrm{eff}} = \rho_{\textrm{inr},s} \mathbf{R}_{ps}$.~The lemma that follows assists in solving problem \eqref{P5} with respect to vector $\bm{w}_1$, when the objective function is given by expression \eqref{SU_rate_high_INR}.
\begin{lemma}
 Considering an interference-limited (high INR) system scenario, optimization problem \eqref{P5} can be approximated by the following problem, the objective of which is a lower bound of the objective of problem \eqref{P5}
\begin{equation}
 \bm{w}_1^{*} = \arg \underset{\bm{w} \in \mathbb{C}^{M \times 1}, \|\bm{w}\| = 1}{\max} \bm{w}^{\He} \tilde{\mathbf{H}}_{\textrm{eff}} \bm{w} + \frac{\bm{w}^{\He} \bar{\mathbf{H}}_{\textrm{eff}} \bm{w}}{\bm{w}^{\He} \mathbf{R}_{\textrm{eff}} \bm{w}},
\tag{P6}
\label{P6}
\end{equation}
where 
\begin{equation}
 \tilde{\mathbf{H}}_{\textrm{eff}} = \kappa_1 \mathbf{H}_{\textrm{eff}}, \hspace{0.1in} \kappa_1 = \frac{f_0(\lambda_{\textrm{max}}(\mathbf{H}_{\textrm{eff}})) - f_0(\lambda_{\textrm{min}}(\mathbf{H}_{\textrm{eff}}))}{\lambda_{\textrm{max}}(\mathbf{H}_{\textrm{eff}}) - \lambda_{\textrm{min}}(\mathbf{H}_{\textrm{eff}})},
\label{tilde_H_eff}
\end{equation}
and
\begin{equation}
 \bar{\mathbf{H}}_{\textrm{eff}} = \mu_1 \mathbf{H}_{\textrm{eff}}, \hspace{0.1in} \mu_1 = \frac{f_1(\lambda_{\textrm{max}}(\mathbf{R}_{\textrm{eff}}^{-1} \mathbf{H}_{\textrm{eff}})) - f_1(\lambda_{\textrm{min}}(\mathbf{R}_{\textrm{eff}}^{-1} \mathbf{H}_{\textrm{eff}}))}{\lambda_{\textrm{max}}(\mathbf{R}_{\textrm{eff}}^{-1} \mathbf{H}_{\textrm{eff}}) - \lambda_{\textrm{min}}(\mathbf{R}_{\textrm{eff}}^{-1} \mathbf{H}_{\textrm{eff}})}.
\label{bar_H_eff}
\end{equation}
Functions $f_0(\cdot)$ and $f_1(\cdot)$ are given by: $f_0(x_1) = \hat{\alpha}_1 \ln(x_1)$ and $f_1(x_2) = \hat{\beta}_1 (\ln(x_2) + \exp(x_2) E_1(x_2))$.
\begin{proof}
 The proof is included in Appendix \ref{proof_approx_prob}.
\end{proof}
\label{approx_prob}
\end{lemma}
Problem \eqref{P6}, i.e., the problem of maximizing the sum of a quadratic form and a Rayleigh quotient over the unit sphere, can be efficiently solved by applying the Trust Region Self Consistent Field (TRSCF) algorithm which was introduced and evaluated in \cite[Algorithm 2]{Zhang2014}.

An interesting sub-case, which is worth investigating, is the case where $\tilde{\mathcal{P}}_d \rightarrow 1, \hspace{0.1in} \mathcal{P}_1 \rightarrow 1$. When the primary system is (almost) always in transmission mode, then, by focusing on an interference-limited system scenario, we obtain an expression reminiscent of \cite[eq. (12)]{FilippouVTC2013} for the SIMO interference channel, which is the following
\begin{equation}
 \mathbb{E}_{|\bm{h}_{ss}}\{\mathcal{R}_1\} \xrightarrow{\mathcal{P}_1 \rightarrow 1, \tilde{\mathcal{P}}_d \rightarrow 1} \frac{\hat{\beta}_1}{\ln(2)} \Bigg(\ln\left(\frac{\bm{w}_{1}^{\He} \mathbf{H}_{\textrm{eff}} \bm{w}_{1}}{\bm{w}_{1}^{\He} \mathbf{R}_{\textrm{eff}} \bm{w}_{1}} \right) + \exp\left(\frac{\bm{w}_{1}^{\He} \mathbf{H}_{\textrm{eff}} \bm{w}_{1}}{\bm{w}_{1}^{\He} \mathbf{R}_{\textrm{eff}} \bm{w}_{1}}\right) E_1\left(\frac{\bm{w}_{1}^{\He} \mathbf{H}_{\textrm{eff}} \bm{w}_{1}}{\bm{w}_{1}^{\He} \mathbf{R}_{\textrm{eff}} \bm{w}_{1}}\right) + \gamma \Bigg).
\label{asymp_regime_obj}
\end{equation}
In such a case, the optimal receive BF can be found as shown in the following proposition.
\begin{proposition}\label{RX_BF_high_act}
For fixed SS parameters, along with a given transmit power level, $P_1^{*}$, which satisfies the constraints of \eqref{P2}, and assuming that $\tilde{\mathcal{P}}_d \rightarrow 1, \mathcal{P}_1 \rightarrow 1$, as well as that the investigated system is interference-limited, the optimal receive BF vector at $\textrm{RX}~s$, in terms of maximizing the conditional (with respect to channel $\bm{h}_{ss}$) average rate of the secondary system, is given by the following expression
\begin{equation}
 \bm{w}_{1}^{*} = \underset{\bm{w}_{1} \in \mathbb{C}^{M \times 1}, \|\bm{w}_{1}\| = 1}{\arg \max} \frac{\bm{w}_{1}^{\He} \left(\mathbf{I}_M + \frac{P_1^{*}}{N_{0,s}} \bm{h}_{ss} \bm{h}_{ss}^{\He}\right) \bm{w}_{1}}{\bm{w}_{1}^{\He} \rho_{\textrm{inr},s} \mathbf{R}_{ps} \bm{w}_{1}},
\label{optimal_RX_BF}
\end{equation}
where $P_1^{*}$ has been obtained in \eqref{PP_3}.~The solution of the latter problem is the eigenvector that corresponds to the dominant eigenvalue of matrix ${\left(\rho_{\textrm{inr},s} \mathbf{R}_{ps}\right)}^{-1} \left(\mathbf{I}_M + \frac{P_1^{*}}{N_{0,s}} \bm{h}_{ss} \bm{h}_{ss}^{\He}\right)$.
\begin{proof}
 The proof is included in Appendix \ref{proof_RX_BF_high_act}.
\end{proof}
\end{proposition}

\subsubsection{Iterative optimization framework}
Having solved separately the SS and BF optimization problems, we propose to approximate the solution to the rate-optimal joint SS and BF design by applying the following iterative algorithm.
\begin{algorithm}[!h]
\caption{Jointly optimizing BF vector $\bm{w}_1$ and SS parameters $\tau$ and $\varepsilon$}\label{alg:iter_algo}
\begin{enumerate}[1]
\item Initialization ($n=0$).~Fix the receive BF scheme such that $\bm{w}_1 = \bm{w}_1^{(0)}$ and increase counter by one.
\item {For the $n$-th iteration, solve problem \eqref{P4} with $\bm{w}_1 = \bm{w}_1^{(n-1)}$ and find values $\tau_n$ and $\varepsilon_n$.}
\item Utilizing values $\tau_n$ and $\varepsilon_n$, solve problem \eqref{P6} and determine BF vector $\bm{w}_1^{(n)}$.
\item Compute the value of the objective $\mathcal{C}_n(\bm{w}_1^{(n)}, \tau_n, \varepsilon_n)$.
\item Increase the counter by one and if $|\mathcal{C}_n - \mathcal{C}_{n-1}| < \xi$, where $n \geq 2$ and $\xi > 0, \xi \in \mathbb{R}$ is an arbitrary small number, stop, otherwise go to Step 2.
\end{enumerate}
\label{Algorithm_1}
\end{algorithm}
\begin{remark}
 Since for $\mathcal{P}_1 \rightarrow 0$, $\bm{w}_1^{*} = \frac{\bm{h}_{ss}}{\|\bm{h}_{ss}\|} = \bm{w}_{\textrm{MRC}}$, while in the case where $\mathcal{P}_1 \rightarrow 1$, the optimal receive BF vector is the DGE of matrices $\mathbf{H}_{\textrm{eff}}$ and $\mathbf{R}_{\textrm{eff}}$, i.e., $\bm{w}_1^{*} = \arg \max_{\|\bm{w}\| = 1} \frac{\bm{w}^{\He} \mathbf{H}_{\textrm{eff}} \bm{w}}{\bm{w}^{\He} \mathbf{R}_{\textrm{eff}} \bm{w}} = \bm{w}_{\textrm{DGE}}$, a heuristic can be exploited in terms of choosing vector $\bm{w}_1^{(0)}$.~For instance, one can use $\bm{w}_1^{(0)} = \bm{w}_{\textrm{MRC}}$ when the primary activity profile is low, otherwise vector $\bm{w}_1^{(0)} = \bm{w}_{\textrm{DGE}}$ can be used.~Such a heuristic can be proved useful in terms of reducing the complexity of Algorithm \ref{alg:iter_algo}.
\end{remark}

In what follows, we focus on the standard interweave and underlay CR systems, and optimization problems, equivalent to \eqref{P2}, are formulated and then solved.
\section{Optimizing design Parameters for Standard CR Systems}
The goal of this section is to derive rate-optimal system designs for interweave and underlay CR systems.~In what follows, we start with the interweave (opportunistic) CR system.
\subsection{Interweave CR system}
Focusing on the interweave CR system, we assume that $\textrm{TX}~s$ transmits with a fixed power level, $P_{\textrm{peak}}$ and the receive BF vector at $\textrm{RX}~s$ is based on the MRC solution\footnote{Regarding the transmit power of $\textrm{TX}~s$ for the interweave case, we choose power level $P_{\textrm{peak}}$, because it is assumed that $\tilde{\mathcal{P}}_d \rightarrow 1$.}.~The rate-optimal design for the interweave CR system boils down to the following problem
\begin{equation}
\begin{aligned}
  & \underset{\tau_{\textrm{int}}, \varepsilon_{\textrm{int}}}{\text{maximize}}
	& & \mathcal{C}_{\textrm{int}} = \frac{\alpha_0(\tau_{\textrm{int}}, \varepsilon_{\textrm{int}})}{\ln(2)} \mathcal{C}_{0,0} + \frac{\beta_0(\tau_{\textrm{int}}, \varepsilon_{\textrm{int}})}{\ln(2)} \mathcal{C}_{0,1} \\
	& \text{subject to} 
	& & \mathcal{P}_{\textrm{out},\textrm{int}} = \tilde{\mathcal{P}}_{\textrm{out}}, \hspace{0.1in} 0 < \tau_{\textrm{int}} \leq T, \hspace{0.1in} \varepsilon_{\textrm{int}} \geq 0,
\end{aligned}
\tag{P7}
\label{P7}
\end{equation}
where the objective is given by the lower bound in \eqref{C_0} and quantities $\alpha_0(\tau_{\textrm{int}}, \varepsilon_{\textrm{int}})$ and $\beta_0(\tau_{\textrm{int}}, \varepsilon_{\textrm{int}})$ have the form of $\alpha_0$ and $\beta_0$, (given in \eqref{alphas_betas}), respectively, by exploiting the new SS parameters, $\tau_{\textrm{int}}$ and $\varepsilon_{\textrm{int}}$.~The following proposition will be useful for solving problem \eqref{P7}.
\begin{proposition}\label{Outage_prob_inter}
 The outage probability of primary communication, for a SIMO interweave CR system is approximated by the expression that follows
\begin{equation}
 \mathcal{P}_{\textrm{out},\textrm{int}} \approx (1 - \mathcal{P}_d) \mathcal{F}(P_{\textrm{peak}}) + \mathcal{P}_d \mathcal{G}, 
\label{P_out_int}
\end{equation}
where
\begin{equation}
 \mathcal{G} = \frac{1}{\prod_{j=1}^M \lambda_j(\mathbf{R}_{pp})} \sum_{j=1}^M \frac{\lambda_j(\mathbf{R}_{pp}) \left(1 - \exp\left(-\frac{\gamma_0}{\rho_{\textrm{snr},p} \lambda_j(\mathbf{R}_{pp})}\right)\right)}{\prod_{k=1, k \neq j}^M \left(\frac{1}{\lambda_k(\mathbf{R}_{pp})} - \frac{1}{\lambda_j(\mathbf{R}_{pp})}\right)},
\label{factor_g}
\end{equation}
and $\rho_{\textrm{snr},p} = \frac{P_p}{N_{0,p}}$ stands for the system Signal-to-Noise Ratio (SNR) observed at $\textrm{RX}~p$.
\begin{proof}
 The proof is included in Appendix \ref{proof_Outage_prob_inter}.
\end{proof}
\end{proposition}
Having derived a closed form approximation for the outage probability of primary communication, one can express the ED threshold, $\varepsilon_{\textrm{int}}$, as a function of sensing time, $\tau_{\textrm{int}}$, after substituting \eqref{P_out_int} to the outage probability constraint of \eqref{P7} and exploiting the closed form approximation for the average detection probability, which is given by \eqref{P_d}.~This expression is the following
\begin{equation}
 \varepsilon_{\textrm{int}} = \delta \left(\frac{\xi_{\textrm{int}}}{\sqrt{\tau_{\textrm{int}} f_s}} + 1\right),
\label{outage_constr_inter}
\end{equation}
where, $\xi_{\textrm{int}} = \mathcal{Q}^{-1}\left(\frac{\tilde{\mathcal{P}}_{\textrm{out}} - \mathcal{F}(P_{\textrm{peak}})}{\mathcal{G} - \mathcal{F}(P_{\textrm{peak}})}\right)$ and $\delta = N_{0,0} \left(1 + \frac{P_p}{N_{0,0}} \sigma_{0}^2\right)$.~Substituting \eqref{outage_constr_inter} to the objective function of \eqref{P7}, one obtains a single variable objective function: $\mathcal{U}(\tau_{\textrm{int}}) = \frac{\alpha_0(\tau_{\textrm{int}})}{\ln(2)} \mathcal{C}_{0,0} + \frac{\beta_0(\tau_{\textrm{int}})}{\ln(2)} \mathcal{C}_{0,1}$.~By applying the second derivative criterion, it can be shown that $\mathcal{U}(\tau_{\textrm{int}})$ is a concave function of its argument, when $\tau_{\textrm{int}} \in (0,T]$, consequently, an optimal $\tau_{\textrm{int}}^{*}$ and a corresponding (by exploiting \eqref{outage_constr_inter}) $\varepsilon_{\textrm{int}}^{*}$ can be found, by applying a convex optimization algorithm.

In what follows, the optimal parameter design problem is formulated and solved for an underlay CR system.
\subsection{Underlay CR System}
Concentrating on the corresponding underlay CR system, the optimization problem, equivalent to \eqref{P2}, that has to be solved, is the following
\begin{equation}
\begin{aligned}
  & \underset{\bm{w}_{\textrm{und}} \in \mathbb{C}^{M \times 1}, P_{\textrm{und}}}{\text{maximize}}
	& & \mathcal{C}_{\textrm{und}} \\
	& \text{subject to} 
	& & \mathcal{P}_{\textrm{out},\textrm{und}} \leq \tilde{\mathcal{P}}_{\textrm{out}}, \hspace{0.1in} \|\bm{w}_{\textrm{und}}\| = 1,
\end{aligned}
\tag{P8}
\label{P8}
\end{equation}
where, $\bm{w}_{\textrm{und}}$ represents the applied receive BF vector at $\textrm{RX}~s$ and $P_{\textrm{und}}$ denotes the transmit power of $\textrm{TX}~s$.~Due to the lack of a SS procedure ($\tau=0$), the conditional average rate of secondary communication is given by the following expression
\begin{equation}
 \mathcal{C}_{\textrm{und}} = \frac{\mathcal{P}_0}{\ln(2)} \mathcal{D}_{1,0}^{\textrm{und}} + \frac{\mathcal{P}_1}{\ln(2)} \mathcal{D}_{1,1}^{\textrm{und}},
\label{avg_rate_under} 
\end{equation}
where quantities $\mathcal{D}_{1,0}^{\textrm{und}}$ and $\mathcal{D}_{1,1}^{\textrm{und}}$ are given by \eqref{D_1_0} and \eqref{D_1_1}, respectively, with $\bm{w}_1 \triangleq \bm{w}_{\textrm{und}}$ and $P_1 \triangleq P_{\textrm{und}}$.
A closed form approximation of the outage probability of primary communication, considering an underlay CR system, denoted as $\mathcal{P}_{\textrm{out},\textrm{und}}$, is given in the following proposition.
\begin{proposition}
 The outage probability of primary communication, for a SIMO underlay CR system is approximated by the following expression
\begin{equation}
 \mathcal{P}_{\textrm{out},\textrm{und}} \approx \mathcal{F}(P_{\textrm{und}}). 
\label{P_out_und_1}
\end{equation}
\begin{proof}
 The outage probability of primary communication is given by
\begin{equation}
 \mathcal{P}_{\textrm{out},\textrm{und}} = Pr\left(\frac{P_p \|\bm{h}_{pp}\|^2}{N_{0,p} + P_{\textrm{und}} |\tilde{\bm{h}}_{pp}^{\He} \bm{h}_{sp}|^2} < \gamma_0\right).
\label{P_out_und_2}
\end{equation}
The latter probability has been approximated in Appendix \ref{proof_Outage_prob}, which concludes the proof.
\end{proof}
\label{propos_Outage_prob_under}
\end{proposition}
Having derived an approximate expression for the outage probability of primary communication, in closed form, problem \eqref{P8} can be efficiently solved.~More specifically, by following the steps of Section \ref{Opt_PP} with $\tilde{\mathcal{P}}_d \rightarrow 1$, the transmit power of $\textrm{TX}~s$ can be determined and by following the steps of Section \ref{opt_BF_fixed_SS}, with $\hat{\alpha}_1 = \mathcal{P}_0$ and $\hat{\beta}_1 = \mathcal{P}_1$, one can find the rate-optimal receive BF vector at $\textrm{RX}~s$.

In the following section, the throughput performance of the designed hybrid CR system is numerically evaluated and compared to the throughput performance achieved by the designed interweave and underlay CR systems.
\section{Numerical Evaluation}
In this section, the throughput performance of the designed hybrid CR system is evaluated and compared to the throughput performance achieved by the equivalent standard interweave and underlay CR systems.~We use Monte Carlo (MC) simulations with 2500 channel realizations, in order to evaluate the performance of the designed CR systems.~An interference-limited system is assumed, the parameters of which are included in Table \ref{Tb:basicParas}.~It should be noted that the values of these parameters remain fixed in the remainder of this section, unless otherwise stated.
\begin{table} [!ht]
\caption{Basic simulation parameters}\label{Tb:basicParas}
\centering
    \begin{tabular}{l|l}
  \hline
  MAC frame size, $T$                                              & 100msec \\
    \hline
  Number of receive antennas, $M$                                  & 4 \\
    \hline
  SINR threshold, $\gamma_0$                                       & 3 dB \\
    \hline
  Sampling frequency, $f_s$                                        & 6 MHz \\
    \hline
  Noise variance, $N_0 = N_{0,0} = N_{0,p} = N_{0,s}$              & 0 dB \\
    \hline
  Antenna correlation factor, $\rho$                               & 0.5 \\
    \hline   
  Power level, $P_{\textrm{peak}}$                                 & 10 dB \\
    \hline
  Power level, $P_p$                                               & 10 dB \\
    \hline
	Variance of $\textrm{TX}~p$-$\textrm{TX}~s$ channel, $\sigma_{0}^2$ & -3dB  \\
	  \hline
	Targeted average detection probability, $\tilde{\mathcal{P}}_d$  & 0.975
\end{tabular}
\end{table}
The exponential antenna correlation model is adopted, as described in \cite{Loyka2001}.~More specifically, considering the $(p,q)$-th element of the covariance matrix of channel $\bm{h}_{mn}, \hspace{0.1in} m,n \in \{p,s\}$, it is taken to be ${[\mathbf{R}_{mn}]}_{(p,q)} = \rho^{|p-q|}, \hspace{0.1in} p,q = 1,\ldots,M, \hspace{0.1in} \rho \in [0,1]$.

\begin{figure}
  \centering
  \includegraphics[scale=0.68]{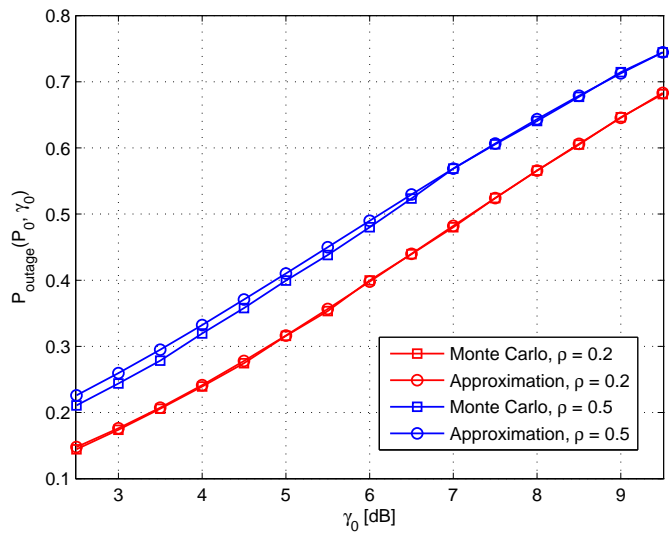}
  \caption{Quality of approximating probability $\mathcal{P}_{\textrm{outage}}(P_0, \gamma_0)$ by value $\mathcal{F}(P_0)$ for different values of $\gamma_0$.}
\label{fig:approx_prob_calF}
\end{figure}

We start with evaluating the quality of approximating $\mathcal{P}_{\textrm{outage}}(P_0, \gamma_0) = Pr\left(\frac{P_p \|\bm{h}_{pp}\|^2}{N_{0,p} + P_0 |\tilde{\bm{h}}_{pp}^{\He} \bm{h}_{sp}|^2} < \gamma_0\right)$ by value $\mathcal{F}(P_0)$ for different values of SINR threshold $\gamma_0$, when $\rho = 0.2$ as well as when $\rho = 0.5$.~As it is evident from Fig.~\ref{fig:approx_prob_calF}, the approximation is satisfactory for the examined range of $\gamma_0$ when $\rho = 0.2$, but also when $\rho = 0.5$.~In the latter case, the approximation quality becomes higher for relatively high values of $\gamma_0$.

\begin{figure}
  \centering
  \includegraphics[scale=0.68]{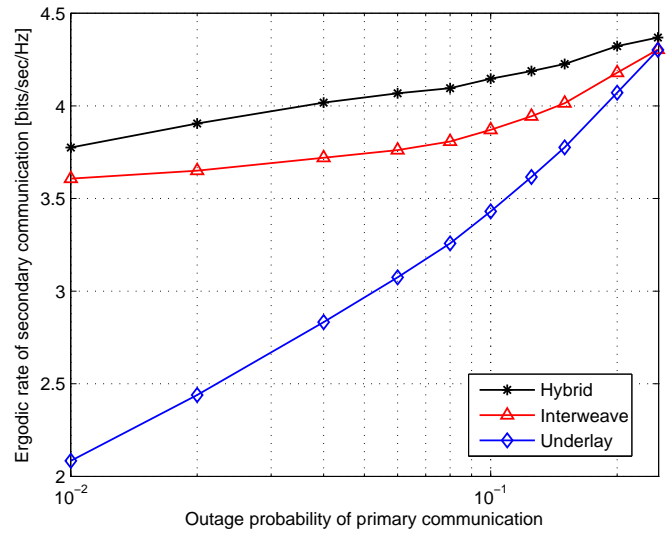}
  \caption{Ergodic rate of $\textrm{RX}~s$ vs. outage probability of primary communication, $\mathcal{P}_1=0.3$.}
\label{fig:Erg_Rate_vs_P_out_low_activity}
\end{figure}

In Fig.~\ref{fig:Erg_Rate_vs_P_out_low_activity}, the average rate of $\textrm{RX}~s$ is depicted as a function of the outage probability of primary communication, when the primary system is in transmission mode for 30$\%$ of the time.~The throughput performance of the optimized hybrid CR system is plotted together with the one achieved by the optimized interweave and underlay CR systems.~One can observe that the performance of the hybrid CR system overcomes the one achieved by the standard CR systems for the whole examined outage probability range.~Also, all three curves are monotonically increasing, which can be explained by the fact that, as the outage probability constraint becomes looser, the secondary system can utilize its available resources primarily with the aim of maximizing its spectral efficiency.~In addition, the average secondary rate, achieved by the interweave system outperforms the one of the underlay system for almost the whole examined outage probability interval.~This happens, because for low primary activity profiles and for the given quality of the SS channel, it is better to sense the existence of spectral ``holes'' in time, in order to then exploit the full potential of the secondary system's resources (i.e., full transmit power).
\begin{figure}
  \centering
  \includegraphics[scale=0.68]{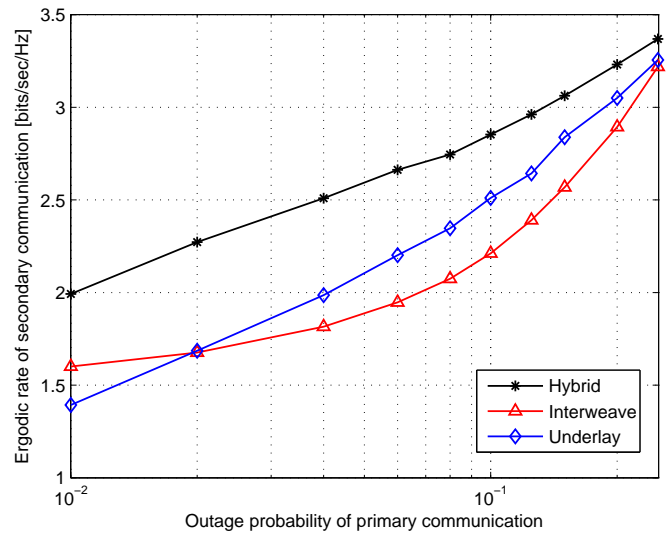}
  \caption{Ergodic rate of $\textrm{RX}~s$ vs. outage probability of primary communication, $\mathcal{P}_1=0.7$.}
\label{fig:Erg_Rate_vs_P_out_high_activity}
\end{figure}

The same performance metric is illustrated in Fig.~\ref{fig:Erg_Rate_vs_P_out_high_activity}, this time for a high activity profile of the PU, i.e., when it is active for 70$\%$ of the time.~In this case, the following observations can be made: \begin{inparaenum}[\itshape a\upshape)] \item The performance of the optimized hybrid system always overcomes the one achieved by the optimized standard CR systems, however, the average secondary rates of all systems are lower than the ones achieved given a low primary activity profile.~This occurs because more interference from the primary system is received by $\textrm{RX}~s$, on average. \item The underlay CR system now outperforms the interweave one for almost the whole investigated outage probability interval.~Such behavioral change can be explained by the fact that, as the primary system transmits more frequently, it is better for the secondary one to exploit the full duration of the MAC frame for DT\end{inparaenum}.
\begin{figure}[!ht]
  \centering
  \includegraphics[scale=0.68]{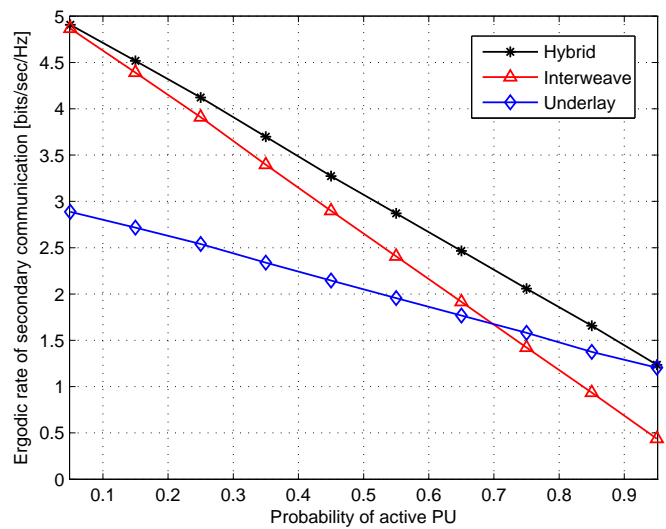}\\
	\vspace{-0.2in}
  \caption{Ergodic rate of $\textrm{RX}~s$ vs. primary activity profile, $\tilde{\mathcal{P}}_{\textrm{out}}=2 \times 10^{-2}$.} \label{fig:Erg_Rate_vs_activity_prof}
\end{figure}

In Fig.~\ref{fig:Erg_Rate_vs_activity_prof}, the achievable average rate of $\textrm{RX}~s$ is depicted for the three investigated systems, as a function of the activity profile of the primary system, when the outage probability of primary communication is equal to 2$\%$.~One can observe that the average throughput of $\textrm{RX}~s$ regarding the hybrid system, balances between two ``extremes'' with respect to the activity profile of the PU.~More specifically, the hybrid CR system behaves similarly to the interweave one, when the PU is idle for most of the time, whereas it approaches the throughput performance of the underlay system, when the PU is active for most of the time.~Also importantly, all three curves are decreasing.~This occurs because, when the primary system is busy for an increased fraction of time, more interference will be received by $\textrm{RX}~s$, on average.
\begin{figure}[!ht]
  \centering
  \includegraphics[scale=0.68]{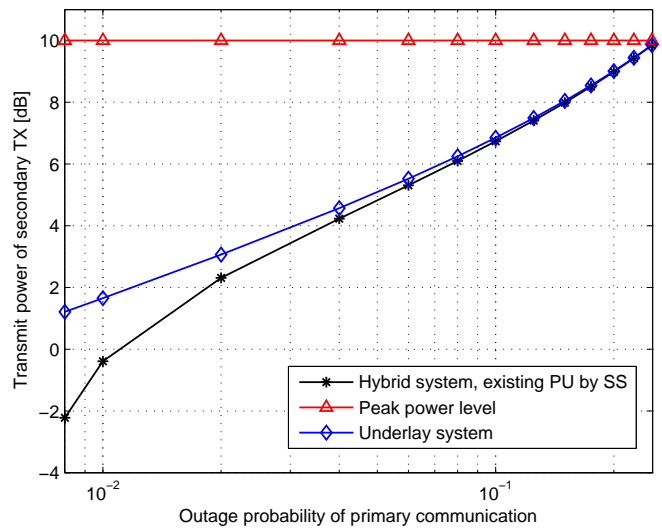}\\
	\vspace{-0.2in}
  \caption{Transmit power of $\textrm{TX}~s$ vs. targeted PU outage probability, $\mathcal{P}_1 = 0.3$.} \label{fig:powers_vs_P_out}
\end{figure}

Finally, in Fig.~\ref{fig:powers_vs_P_out}, the transmit power levels $P_1$ of the hybrid CR system and $P_{\textrm{und}}$ of the equivalent underlay CR system are shown as a function of the targeted outage probability of the PU, when the primary system is active for 30$\%$ of the time.~It is clearly shown that the transmit power of $\textrm{TX}~s$ increases as the PU becomes more tolerant to interference.~It is also observed that when the outage probability constraint becomes very loose, i.e., when $\tilde{\mathcal{P}}_{\textrm{out}}$ is about 25$\%$, both the hybrid and the underlay systems exploit that flexibility in order to transmit with full power. 

\section{Conclusions}
In this paper, the uplink of an interference-limited, hybrid interweave/underlay CR system has been studied.~Correlated Rayleigh fading has been assumed for the involved direct and interference channels.~A realistic CSIR scenario has been examined, according to which each RX has mere access to the instantaneous link of its assigned terminal, along with covariance information regarding the global uplink channel.~Novel closed form approximations, describing the outage probability of the primary system have been derived, considering the hybrid CR system as well as the standard interweave and underlay CR systems.~Exploiting these expressions, a new optimization problem that consists in jointly optimizing the SS parameters and the applied receive BF scheme, towards maximizing the average secondary rate, subject to an outage probability-based constraint for primary communication, has been formulated and solved for all the examined systems.~It has been numerically shown that the optimized hybrid CR system outperforms the equivalent, optimized interweave and underlay CR systems, in terms of spectral efficiency.~Interesting extensions can be made regarding the existence of multiple secondary terminals.

\begin{appendices}
\section{Proof of Proposition \ref{Outage_prob}}\label{proof_Outage_prob}
The outage probability of primary communication is given by the following expression
\begin{equation}
 \mathcal{P}_{\textrm{out}} = \mathcal{P}_{\textrm{out},0} + \mathcal{P}_{\textrm{out},1},
\label{propos_1_1}
\end{equation}
where term $\mathcal{P}_{\textrm{out},k}$ corresponds to the occurrence of event $\hat{\mathcal{H}}_k, \hspace{0.1in} k=0,1$.~Hence, for the first term of \eqref{propos_1_1}, one obtains
\begin{equation}
 \mathcal{P}_{\textrm{out},0} = \left(1 - \mathcal{P}_d\right) Pr\left(\frac{P_p \|\bm{h}_{pp}\|^2}{N_{0,p} + P_0 |\tilde{\bm{h}}_{pp}^{\He} \bm{h}_{sp}|^2} < \gamma_0\right).
\label{term_1_P_out}
\end{equation}
Let us start by defining the RVs $X = \tilde{\bm{h}}_{pp}^{\He} \bm{h}_{sp}$ and $Y = |X|^2$.~ Assuming that vector $\bm{h}_{pp}$ is given and that $\bm{h}_{sp} = \mathbf{R}_{sp}^{\frac{1}{2}} \bm{h}_{sp,w}$, where $\bm{h}_{sp,w} \sim \mathcal{CN}(\bm{0}_{M}, \mathbf{I}_M)$, one can write $X$ as: $X = \tilde{\bm{h}}_{pp}^{\He} \mathbf{R}_{sp}^{\frac{1}{2}} \bm{h}_{sp,w}$.~It then holds that, given $\bm{h}_{pp}$, $X$ is a complex normal RV with zero mean and variance $\sigma_{X}^2 = {\|\tilde{\bm{h}}_{pp}^{\He} \mathbf{R}_{sp}^{\frac{1}{2}}\|}^2 = \frac{\bm{h}_{pp}^{\He} \mathbf{R}_{sp} \bm{h}_{pp}}{{\|\bm{h}_{pp}\|}^2}$, therefore, variance $\sigma_{X}^2$ is a ratio of quadratic forms.~Hence, given $\bm{h}_{pp}$, $Y$ is an exponential RV with mean value equal to $\sigma_{X}^2$, i.e., it has a Probability Density Function (PDF) of the form
\begin{equation}
 f_{Y}(y|\bm{h}_{pp}) = \frac{1}{\sigma_{X}^2} \exp\left(-\frac{y}{\sigma_{X}^2}\right),
\label{pdf_Y}
\end{equation}
and a Cumulative Distribution Function (CDF), given by
\begin{equation}
 F_{Y}(y|\bm{h}_{pp}) = 1 - \exp\left(-\frac{y}{\sigma_{X}^2}\right).
\label{cdf_Y}
\end{equation}
Conditioned on the knowledge of $\bm{h}_{pp}$, one can write \eqref{term_1_P_out} as
\begin{equation}
\begin{aligned}
 \mathcal{P}_{\textrm{out},0 |\bm{h}_{pp}} &= \left(1 - \mathcal{P}_d\right) Pr\left(Y > \frac{\|\bm{h}_{pp}\|^2 P_p}{\gamma_0 P_0} - \frac{N_{0,p}}{P_0} | \bm{h}_{pp} \right) \\
&= \left(1 - \mathcal{P}_d\right) \exp\left(- \frac{\|\bm{h}_{pp}\|^2 P_p}{\gamma_0 P_0 \sigma_{X}^2} + \frac{N_{0,p}}{P_0 \sigma_{X}^2}\right).
\end{aligned}
\label{P_out_0_2}
\end{equation}
Assuming that covariance matrix $\mathbf{R}_{pp}$ has $M$ distinct eigenvalues, RV $Z = \|\bm{h}_{pp}\|^2$ is distributed with PDF given by \cite[eq. (14)]{Mallik2004}
\begin{equation}
 f_{Z}(z) = \frac{1}{\prod_{j=1}^M \lambda_j(\mathbf{R}_{pp})} \sum_{j=1}^M \frac{\exp\left(-\frac{z}{\lambda_j(\mathbf{R}_{pp})}\right)}{\prod_{k=1, k \neq j}^M \left(\frac{1}{\lambda_k(\mathbf{R}_{pp})} - \frac{1}{\lambda_j(\mathbf{R}_{pp})}\right)}, \hspace{0.1in} z \geq 0.
\label{pdf_Z1}
\end{equation}
As a result, probability $\mathcal{P}_{\textrm{out},0}$ can be approximated by the following expression
\begin{equation}
 \mathcal{P}_{\textrm{out},0} \approx \left(1 - \mathcal{P}_d\right) \int_0^{\infty} \exp\left(-\frac{P_p z}{\gamma_0 P_0 \bar{\lambda}} + \frac{N_{0,p}}{P_0 \bar{\lambda}}\right) f_Z(z) dz,
\label{P_out_0_3}
\end{equation}
where $\bar{\lambda} = \mathbb{E}\{\sigma_{X}^2\}$.~The latter expectation can be computed in closed form by exploiting \cite[Lemma 3]{FilippouStatCoord2015}, with matrices $\mathbf{A} = \mathbf{R}_{pp}$ and $\mathbf{B} = \mathbf{R}_{pp}^{\frac{1}{2}} \mathbf{R}_{sp} \mathbf{R}_{pp}^{\frac{1}{2}}$.~Consequently, term $\mathcal{P}_{\textrm{out},0}$ becomes
\begin{equation}
 \mathcal{P}_{\textrm{out},0} \approx \left(1 - \mathcal{P}_d\right) \mathcal{F}(P_0),
\label{P_out_0_4}
\end{equation}
where function $\mathcal{F}(x)$ is given in \eqref{function_f}.~Following a similar analysis, probability $\mathcal{P}_{\textrm{out},1}$ is given by the following expression
\begin{equation}
 \mathcal{P}_{\textrm{out},1} \approx \mathcal{P}_d \mathcal{F}(P_1),
\label{P_out_1}
\end{equation}
which completes the proof.

\section{Proof of Lemma \ref{SU_rate_interweave}}\label{proof_SU_rate_interweave}
In the occurrence of event $\hat{\mathcal{H}}_0$, the average secondary rate, conditioned on the instantaneous knowledge of channel $\bm{h}_{ss}$ at $\textrm{RX}~s$, is given by
\begin{equation}
 \mathbb{E}_{|\bm{h}_{ss}}\left\{\mathcal{R}_0\right\} = \alpha_0 \log_2\left(1 + \frac{P_{\textrm{peak}} \|\bm{h}_{ss}\|^2}{N_{0,s}}\right) + \beta_0 \mathbb{E}_{|\bm{h}_{ss}}\left\{\log_2\left(1 + \frac{P_{\textrm{peak}} \|\bm{h}_{ss}\|^2}{N_{0,s} + u_0}\right)\right\},
\label{SU_rate_0}
\end{equation}
where $u_0 = P_p |\tilde{\bm{h}}_{ss}^{\He} \bm{h}_{ps}|^2$.~For the expectation appearing in the second term of \eqref{SU_rate_0}, by applying Jensen's inequality with respect to channel $\bm{h}_{ps}$, we obtain \cite{cover2012elements}
\begin{equation}
\begin{aligned}
 \mathbb{E}_{|\bm{h}_{ss}}\left\{\log_2\left(1 + \frac{P_{\textrm{peak}} \|\bm{h}_{ss}\|^2}{N_{0,s} + u_0}\right)\right\} &\geq \log_2\left(1 + \frac{P_{\textrm{peak}} \|\bm{h}_{ss}\|^2}{N_{0,s} + \mathbb{E}_{\bm{h}_{ps}}\{u_0\}}\right) \\
&= \log_2\left(1 + \frac{P_{\textrm{peak}} \|\bm{h}_{ss}\|^2}{N_{0,s} + P_p \frac{\bm{h}_{ss}^{\He} \mathbf{R}_{ps} \bm{h}_{ss}}{\|\bm{h}_{ss}\|^2}}\right),
\end{aligned}
\label{SU_rate_0_2}
\end{equation}
which completes the proof.
\section{Proof of Lemma \ref{SU_rate_underlay}}\label{proof_SU_rate_underlay}
The achievable average rate of secondary communication, conditioned on the instantaneous knowledge of channel $\bm{h}_{ss}$ at $\textrm{RX}~s$ and given that event $\hat{\mathcal{H}}_1$ has occurred, is given by the following expression
\begin{equation}
 \mathbb{E}_{|\bm{h}_{ss}}\left\{\mathcal{R}_1\right\} = \alpha_1 \log_2\left(1 + \frac{P_1 |\bm{w}_{1}^{\He} \bm{h}_{ss}|^2}{N_{0,s}}\right) + \beta_1 \mathbb{E}_{|\bm{h}_{ss}}\left\{\log_2\left(1 + \frac{P_1 |\bm{w}_{1}^{\He} \bm{h}_{ss}|^2}{N_{0,s} + u_1}\right)\right\},
\label{SU_rate_1}
\end{equation}
where $u_1 = P_p |\bm{w}_{1}^{\He} \bm{h}_{ps}|^2$.
Considering the second term of \eqref{SU_rate_1}, we get the following expression
\begin{equation}
 \mathbb{E}_{|\bm{h}_{ss}}\left\{\log_2\left(1 + \frac{P_1 |\bm{w}_{1}^{\He} \bm{h}_{ss}|^2}{N_{0,s} + u_1}\right)\right\} = \mathbb{E}_{|\bm{h}_{ss}}\left\{\log_2\left(1 + \frac{\frac{P_1}{N_{0,s}} |\bm{w}_{1}^{\He} \bm{h}_{ss}|^2}{1 + Y_1}\right)\right\},
\label{SU_rate_1_2}
\end{equation}
where, $Y_1 = \rho_{\textrm{inr},s} |\bm{w}_{1}^{\He} \bm{h}_{ps}|^2$ and BF vector $\bm{w}_{1}$ is independent of $\bm{h}_{ps}$, since no instantaneous knowledge of $\bm{h}_{ps}$ is presumed.~RV $Y_1$ can be written the following way
\begin{equation}
 Y_1 = \rho_{\textrm{inr},s} |\bm{w}_{1}^{\He} \mathbf{R}_{ps}^{\frac{1}{2}} \bm{h}_{ps,w}|^2,
\label{RV_Y_1}
\end{equation}
where $\bm{h}_{ps,w} \sim \mathcal{CN}(\bm{0}_{M}, \mathbf{I}_M)$. 

It is, thus, easy to confirm that $Y_1$ is an exponentially distributed RV, and its PDF is given by: $f_{Y_1}(y_1) = \frac{1}{\rho_{\textrm{inr},s} \bm{w}_1^{\He} \mathbf{R}_{ps} \bm{w}_1} \exp\left(-\frac{y_1}{\rho_{\textrm{inr},s} \bm{w}_1^{\He} \mathbf{R}_{ps} \bm{w}_1}\right)$.~As a result, for the expectation in \eqref{SU_rate_1_2}, one obtains
\begin{equation}
\begin{aligned}
 \mathbb{E}_{|\bm{h}_{ss}}\left\{\log_2\left(1 + \frac{\frac{P_1}{N_{0,s}} |\bm{w}_{1}^{\He} \bm{h}_{ss}|^2}{1 + Y_1}\right)\right\} &= \mathbb{E}_{|\bm{h}_{ss}}\left\{\log_2\left(1 + Y_1 + \frac{P_1}{N_{0,s}} |\bm{w}_{1}^{\He} \bm{h}_{ss}|^2\right)\right\} \\
&- \mathbb{E}_{|\bm{h}_{ss}}\left\{\log_2\left(1 + Y_1\right)\right\}.
\label{SU_rate_1_3}
\end{aligned}
\end{equation}
For the first term of \eqref{SU_rate_1_3}, by exploiting \cite[eq. (4.337.1)]{Gradshteyn2007}, we obtain the following expression
\begin{equation}
\begin{aligned}
 &\mathbb{E}_{|\bm{h}_{ss}}\left\{\log_2\left(1 + Y_1 + \frac{P_1}{N_{0,s}} |\bm{w}_{1}^{\He} \bm{h}_{ss}|^2\right)\right\} = \log_2\left(1 + \frac{P_1}{N_{0,s}} |\bm{w}_{1}^{\He} \bm{h}_{ss}|^2 \right) \\
&+ \frac{1}{\ln(2)} \exp\left(\frac{\bm{w}_{1}^{\He} \left(\mathbf{I}_M + \frac{P_1}{N_{0,s}} \bm{h}_{ss} \bm{h}_{ss}^{\He}\right) \bm{w}_{1}}{\bm{w}_{1}^{\He} \rho_{\textrm{inr},s} \mathbf{R}_{ps} \bm{w}_{1}}\right) E_1\left(\frac{\bm{w}_{1}^{\He} \left(\mathbf{I}_M + \frac{P_1}{N_{0,s}} \bm{h}_{ss} \bm{h}_{ss}^{\He}\right) \bm{w}_{1}}{\bm{w}_{1}^{\He} \rho_{\textrm{inr},s} \mathbf{R}_{ps} \bm{w}_{1}}\right).
\label{SU_rate_1_4}
\end{aligned}
\end{equation}
Also, exploiting \cite[eq. (4.337.2)]{Gradshteyn2007}, one can derive the second term of \eqref{SU_rate_1_3} as follows
\begin{equation}
 \mathbb{E}_{|\bm{h}_{ss}}\left\{\log_2\left(1 + Y_1\right)\right\} = \frac{1}{\ln(2)} \exp\left(\frac{1}{\bm{w}_{1}^{\He} \rho_{\textrm{inr},s} \mathbf{R}_{ps} \bm{w}_{1}}\right) E_1\left(\frac{1}{\bm{w}_{1}^{\He} \rho_{\textrm{inr},s} \mathbf{R}_{ps} \bm{w}_{1}}\right).
\label{SU_rate_1_5}
\end{equation}
Substituting expressions \eqref{SU_rate_1_3} - \eqref{SU_rate_1_5} to \eqref{SU_rate_1}, expression \eqref{SU_rate_under_final} is obtained, which completes the proof.
\section{Proof of Lemma \ref{Concave_SS}}\label{proof_Concave_SS}
The resulting single-variable objective function of \eqref{P4} is expressed as
\begin{equation}
 \tilde{\mathcal{C}}(\tau, \varepsilon(\tau)) = \alpha_0 \mathcal{C}_{0,0} + \beta_0 \mathcal{C}_{0,1} + \alpha_1 \hat{\mathcal{D}}_{1,0} + \beta_1 \hat{\mathcal{D}}_{1,1},
\label{SS_proof_1}
\end{equation}
where $\alpha_i, \beta_i, i=0,1$ have been defined in \eqref{alphas_betas}.~Taking the derivative of \eqref{SS_proof_1}, with respect to $\tau$ and letting $\delta = N_{0,0}\left(1 + \frac{P_p}{N_{0,0}} \sigma_{0}^2\right)$ and $\xi = \mathcal{Q}^{-1}(\tilde{\mathcal{P}}_d)$, we have
\begin{equation}
\begin{aligned}
 &\frac{\partial \tilde{\mathcal{C}}(\tau, \varepsilon(\tau))}{\partial \tau} = -\frac{\mathcal{P}_0}{T}(\mathcal{C}_{0,0} + \mathcal{C}_{0,1}) - \frac{\mathcal{P}_0}{T} (\hat{\mathcal{D}}_{1,0} - \mathcal{C}_{0,0}) \mathcal{Q}\left(\sqrt{\tau f_s} \left(\frac{\delta}{N_{0,0}} - 1\right) + \frac{\delta \xi}{N_{0,0}} \right) \\
&- \frac{(T-\tau) \mathcal{P}_0 (\hat{\mathcal{D}}_{1,0} - \mathcal{C}_{0,0})}{T \sqrt{2 \pi}} \exp\left(-\frac{1}{2}{\left(\sqrt{\tau f_s} \left(\frac{\delta}{N_{0,0}} - 1\right) + \frac{\delta \xi}{N_{0,0}}\right)}^2 \right) \left(\frac{\delta}{N_{0,0}} - 1\right) \frac{f_s}{2 \sqrt{\tau f_s}} \\
&- \frac{1}{T} \mathcal{Q}(\xi) \mathcal{P}_1 (\hat{\mathcal{D}}_{1,1} - \mathcal{C}_{0,1}).
\end{aligned}
\label{SS_proof_2}
\end{equation}
Taking now the derivative of \eqref{SS_proof_2} with respect to $\tau$, one obtains
\small
\begin{equation}
\begin{aligned}
 &\frac{\partial^2 \tilde{\mathcal{C}}(\tau, \varepsilon(\tau))}{\partial \tau^2} = \frac{\mathcal{P}_0 (\hat{\mathcal{D}}_{1,0} - \mathcal{C}_{0,0})}{T \sqrt{2 \pi}} \exp\left(-\frac{1}{2}{\left(\sqrt{\tau f_s} \left(\frac{\delta}{N_{0,0}} - 1\right) + \frac{\delta \xi}{N_{0,0}}\right)}^2 \right) \left(\frac{\delta}{N_{0,0}} - 1\right) \frac{f_s}{\sqrt{\tau f_s}} \\
&+ \frac{(T-\tau) \mathcal{P}_0 (\hat{\mathcal{D}}_{1,0} - \mathcal{C}_{0,0})}{T \sqrt{2 \pi}} \left(\sqrt{\tau f_s} \left(\frac{\delta}{N_{0,0}} - 1\right) + \frac{\delta \xi}{N_{0,0}}\right) {\left(\frac{\delta}{N_{0,0}} - 1\right)}^2 \frac{f_s^2}{4 \tau f_s} \exp\left(-\frac{1}{2}{\left(\sqrt{\tau f_s} \left(\frac{\delta}{N_{0,0}} - 1\right) + \frac{\delta \xi}{N_{0,0}}\right)}^2 \right) \\
&+ \frac{(T-\tau) \mathcal{P}_0 (\hat{\mathcal{D}}_{1,0} - \mathcal{C}_{0,0})}{4 T \sqrt{2 \pi}} \exp\left(-\frac{1}{2}{\left(\sqrt{\tau f_s} \left(\frac{\delta}{N_{0,0}} - 1\right) + \frac{\delta \xi}{N_{0,0}}\right)}^2 \right) \left(\frac{\delta}{N_{0,0}} - 1\right) f_s {(\tau f_s)}^{-\frac{3}{2}}.
\end{aligned}
\label{SS_proof_3}
\end{equation} 
\normalsize
In the above expression we have $\frac{\delta}{N_{0,0}} - 1 = \frac{P_p \sigma_{0}^2}{N_{0,0}} > 0$.~Also, $\hat{\mathcal{D}}_{1,0} - \mathcal{C}_{0,0} < 0$, since it holds that $P_1^{*} \leq P_{\textrm{peak}}$ and $|\hat{\bm{w}}_{1}^{\He} \bm{h}_{ss}|^2 \leq \|\bm{h}_{ss}\|^2$.~As a result, $\frac{\partial^2 \tilde{\mathcal{C}}(\tau, \varepsilon(\tau))}{\partial \tau^2} < 0$ and, thus, according to the second derivative criterion function, $\tilde{\mathcal{C}}(\tau, \varepsilon(\tau))$, is concave when $\tau \in (0,T]$, which completes the proof.
\section{Proof of Lemma \ref{approx_prob}}\label{proof_approx_prob}
We define variables $x_1 \triangleq \bm{w}_1^{\He} \mathbf{H}_{\textrm{eff}} \bm{w}_1$, $x_2 \triangleq \frac{\bm{w}_1^{\He} \mathbf{H}_{\textrm{eff}} \bm{w}_1}{\bm{w}_1^{\He} \mathbf{R}_{\textrm{eff}} \bm{w}_1}$ and function $f(x_1, x_2) = f_0(x_1) + f_1(x_2)$, where $f_0(x_1) = \hat{\alpha}_1 \ln(x_1)$ and $f_1(x_2) = \hat{\beta}_1 (\ln(x_2) + \exp(x_2) E_1(x_2))$.

One can easily observe that: \begin{inparaenum}[(i)] \item function $f_0(x_1)$ is defined for $x_1 \in \mathcal{A}_0 \triangleq [\lambda_{\textrm{min}}(\mathbf{H}_{\textrm{eff}}), \lambda_{\textrm{max}}(\mathbf{H}_{\textrm{eff}})]$ and it is concave in the same interval and \item function $f_1(x_2)$ is defined for $x_2 \in \mathcal{A}_1 \triangleq [\lambda_{\textrm{min}}(\mathbf{R}_{\textrm{eff}}^{-1}\mathbf{H}_{\textrm{eff}}), \lambda_{\textrm{max}}(\mathbf{R}_{\textrm{eff}}^{-1}\mathbf{H}_{\textrm{eff}})]$ and it is concave in the same interval. \end{inparaenum} Since the two single-variable functions are concave within their domains, it can be concluded that
\begin{itemize}
 \item $\forall x_1 \in \mathcal{A}_0, f_0(x_1) \geq z_0 = \kappa_1 x_1 + \kappa_2$, where line $z_0$ is defined by points $(\lambda_{\textrm{min}}(\mathbf{H}_{\textrm{eff}}), f_0(\lambda_{\textrm{min}}(\mathbf{H}_{\textrm{eff}})))$ and $(\lambda_{\textrm{max}}(\mathbf{H}_{\textrm{eff}}), f_0(\lambda_{\textrm{max}}(\mathbf{H}_{\textrm{eff}})))$ and 

 \item $\forall x_2 \in \mathcal{A}_1, f_1(x_2) \geq z_1 = \mu_1 x_2 + \mu_2$, where line $z_1$ is defined by points $(\lambda_{\textrm{min}}(\mathbf{R}_{\textrm{eff}}^{-1}\mathbf{H}_{\textrm{eff}}), \\ f_1(\lambda_{\textrm{min}}(\mathbf{R}_{\textrm{eff}}^{-1}\mathbf{H}_{\textrm{eff}})))$ and $(\lambda_{\textrm{max}}(\mathbf{R}_{\textrm{eff}}^{-1}\mathbf{H}_{\textrm{eff}}), f_1(\lambda_{\textrm{max}}(\mathbf{R}_{\textrm{eff}}^{-1}\mathbf{H}_{\textrm{eff}})))$.
\end{itemize}
As a result, instead of solving optimization problem 
\begin{equation}
 \bm{w}_1^{*} = \arg \underset{\|\bm{w}\| = 1}{\max} f_0(\bm{w}^{\He} \mathbf{H}_{\textrm{eff}} \bm{w}) + f_1\left(\frac{\bm{w}^{\He} \mathbf{H}_{\textrm{eff}} \bm{w}}{\bm{w}^{\He} \mathbf{R}_{\textrm{eff}} \bm{w}}\right),
\tag{$P^{\prime}$}
\label{initial_problem}
\end{equation}
for fixed $\tau = \hat{\tau}$ and $\varepsilon = \hat{\varepsilon}$, an approximated version of it can be solved, where the new objective is a lower bound of the objective of problem \eqref{initial_problem}.~This completes the proof.

\section{Proof of Proposition \ref{RX_BF_high_act}}\label{proof_RX_BF_high_act}
The optimal receive BF vector is obtained by solving the following optimization problem at $\textrm{RX}~s$:
\begin{equation}
\bm{w}_{1}^{*} = \underset{\bm{w}_{1} \in \mathbb{C}^{M \times 1}, \|\bm{w}_{1}\| = 1}{\arg \max}\mathbb{E}_{|\bm{h}_{ss}}\{\mathcal{R}_{1}\},
\tag{P}
\label{P}
\end{equation}
where an approximation of the objective for the investigated regime is given by \eqref{approx_high_INR}.

One can rewrite $\mathbb{E}_{|\bm{h}_{ss}}\{\mathcal{R}_{1}\}$ as
\begin{equation}
 \mathbb{E}_{|\bm{h}_{ss}}\{\mathcal{R}_{1}\} = \mathcal{V}(\mu_{\bm{w}_1}),\;\textrm{where}\;\;\; 
\mu_{\bm{w}_1} = \frac{\bm{w}_{1}^{\He} \left(\mathbf{I}_M + \frac{P_1^{*}}{N_{0,s}} \bm{h}_{ss} \bm{h}_{ss}^{\He}\right) \bm{w}_{1}}{\bm{w}_{1}^{\He} \rho_{\textrm{inr},s} \mathbf{R}_{ps} \bm{w}_{1}}.
\end{equation}
Focusing on the fact that $\tilde{\mathcal{P}}_d \rightarrow 1, \hspace{0.1in} \mathcal{P}_1 \rightarrow 1$, function $\mathcal{V}(\cdot)$ is defined as
\begin{equation}
\mathcal{V}(\mu_{\bm{w}_1})= \frac{\hat{\beta_1}}{\ln(2)}\left(\ln\left(\mu_{\bm{w}_1}\right)  + \exp(\mu_{\bm{w}_1}) E_{1}\left(\mu_{\bm{w}_1}\right) + \gamma\right).
\end{equation}
By differentiating $\mathcal{V}(\mu_{\bm{w}_1})$ and using \cite[eq. (5.1.26)]{Abramowitz}, one can prove that $\mathcal{V}(\mu_{\bm{w}_1})$ is an increasing function of $\mu_{\bm{w}_1}$.~Consequently, the optimization problem \eqref{P} is equivalent to the Rayleigh - Ritz quotient maximization problem 
\begin{equation}\label{Eq:OptimalBFProblem1}
\bm{w}_{1}^{*} = \underset{\bm{w}_{1} \in \mathbb{C}^{M \times 1}, \|\bm{w}_{1}\| = 1}{\arg \max}\mu_{\bm{w}_1}.
\end{equation}
By setting the derivative of $\mu_{\bm{w}_1}$, with respect to $\bm{w}_1$, equal to zero, it can be found that the optimal BF vector is the one satisfying the equality
\begin{equation}\label{Eq:BFSolution}
\left(\mathbf{I}_M + \frac{P_1^{*}}{N_{0,s}} \bm{h}_{ss}\bm{h}_{ss}^{\He}\right) \bm{w}_{1}^{*} = \mu_{\bm{w}_1} \rho_{\textrm{inr,s}} \mathbf{R}_{ps}\bm{w}_{1}^{*}.
\end{equation}
As a result, by inspecting \eqref{Eq:BFSolution}, one can conclude that the optimal BF vector for $\textrm{RX}~s$ is the dominant generalized eigenvector (DGE) of matrix pair $\left( \left(\mathbf{I}_M + \frac{P_1^{*}}{N_{0,s}} \bm{h}_{ss}\bm{h}_{ss}^{\He}\right), \rho_{\textrm{inr,s}} \mathbf{R}_{ps}\right)$.

\section{Proof of Proposition \ref{Outage_prob_inter}}\label{proof_Outage_prob_inter}
Focusing on an interweave CR system, the outage probability of primary communication is defined as follows
\begin{equation}
 \mathcal{P}_{\textrm{out},\textrm{int}} = \underbrace{\left(1 - \mathcal{P}_d\right) Pr\left(\frac{P_p \|\bm{h}_{pp}\|^2}{N_{0,p} + P_{\textrm{peak}} |\tilde{\bm{h}}_{pp}^{\He} \bm{h}_{sp}|^2} < \gamma_0\right)}_{\mathcal{P}_{\textrm{out},\textrm{int},1}} + \underbrace{\mathcal{P}_d Pr\left(\frac{P_p \|\bm{h}_{pp}\|^2}{N_{0,p}} < \gamma_0\right)}_{\mathcal{P}_{\textrm{out},\textrm{int},2}}. 
\label{outage_int_proof_1}
\end{equation}
Exploiting the proof steps of Appendix \ref{proof_Outage_prob}, the first term of \eqref{outage_int_proof_1}, is given by
\begin{equation}
 \mathcal{P}_{\textrm{out},\textrm{int},1} \approx (1 - \mathcal{P}_d) \mathcal{F}(P_{\textrm{peak}}). 
\label{outage_int_proof_2}
\end{equation}
The second term of \eqref{outage_int_proof_1}, is given by the following expression
\begin{equation}
 \mathcal{P}_{\textrm{out},\textrm{int},2} = \mathcal{P}_d Pr\left(\rho_{\textrm{snr},p} \|\bm{h}_{pp}\|^2 < \gamma_0\right).
\label{outage_int_proof_3}
\end{equation}
Since the PDF of RV $Z_1 = \rho_{\textrm{snr},p} \|\bm{h}_{pp}\|^2$, is known, we obtain
\begin{equation}
 \mathcal{P}_{\textrm{out},\textrm{int},2} = \mathcal{P}_d \int_0^{\gamma_0} f_{Z_1}(z_1) dz_1 = \mathcal{P}_d \mathcal{G},
\label{outage_int_proof_4}
\end{equation}
where $\mathcal{G}$ is given in \eqref{factor_g}.~This completes the proof.
\end{appendices}

\ifCLASSOPTIONcaptionsoff
  \newpage
\fi

\bibliography{bib/allCitations}
\bibliographystyle{IEEEtran}

\end{document}